\documentclass[aip,pof,reprint,a4paper,amsmath,amssymb,nofootinbib]{revtex4-2}

\usepackage[T1]{fontenc}
\usepackage[utf8]{inputenc}
\usepackage{newtxtext,newtxmath}
\usepackage{graphicx}
\usepackage{bm}
\usepackage{microtype}
\usepackage{xcolor}
\usepackage[colorlinks=true,linkcolor=blue!50!black,citecolor=blue!50!black,urlcolor=blue!50!black]{hyperref}

\makeatletter
\@booleantrue\twocolumn@sw
\set@page@grid
\renewcommand\frontmatter@abstractwidth{\textwidth}
\def\adjust@abstractwidth{\parindent\z@\advance\leftskip.5in\relax\advance\rightskip.5in\relax\@totalleftmargin\leftskip\@afterheading\@afterindentfalse}
\def\frontmatter@@indent{\skip@\@flushglue\@flushglue\z@ plus0\textwidth\raggedright\advance\leftskip\z@\advance\rightskip\z@\parindent\z@}
\let\orig@fmfootnotetext\frontmatter@footnotetext
\long\def\frontmatter@footnotetext#1{\orig@fmfootnotetext{#1\parfillskip 0pt plus 1fil\relax}}
\makeatother

\usepackage{placeins}
\usepackage{float}

\graphicspath{{figures_v8/}}

\newcommand{\Uinf}{U_\infty}
\newcommand{\nueff}{\nu_\mathrm{eff}}
\newcommand{\xsep}{x_\mathrm{sep}}
\newcommand{\dmin}{d_\mathrm{min}}
\newcommand{\ours}{GP-FNO}
\newcommand{\HoU}{0.50}

\newcommand{\HoP}{1.9}
\newcommand{\HoNu}{43.0}

\newcommand{\HoCp}{2.7}

\newcommand{\HoCl}{1.1}
\newcommand{\HoCf}{6.2}
\newcommand{\NewU}{0.6}
\newcommand{\NewP}{5.9}
\newcommand{\NewCp}{5.3}
\newcommand{\NewCl}{3.2}
\newcommand{\NewCf}{12.0}

\newcommand{\HoMemo}{12.8}
\newcommand{\HoClMax}{6.1}
\newcommand{\GapU}{0.50}
\newcommand{\GapP}{5.3}
\newcommand{\GapCp}{3.7}
\newcommand{\GapCl}{2.8}
\newcommand{\GapCf}{15.1}
\newcommand{\GapNu}{19.5}

\newcommand{\CondTrN}{64}

\newcommand{\CondUnN}{72}

\newcommand{\Nnew}{136}

\newcommand{\LiftMedAbs}{0.026}
\newcommand{\StN}{44}
\newcommand{\OsN}{92}
\newcommand{\StCp}{4.0}
\newcommand{\OsCp}{6.1}
\newcommand{\CfTarget}{10.7}
\newcommand{\CfTargetLo}{6.7}
\newcommand{\CfTargetHi}{40.3}

\newcommand{\LocP}{7.1}

\newcommand{\LocCp}{2.1}

\newcommand{\LocNewCp}{4.9}

\newcommand{\LocNewCf}{14.4}

\newcommand{\LocGf}{5.1}

\newcommand{\NfCp}{2.5}

\newcommand{\NfCf}{7.7}

\newcommand{\NwCp}{2.2}
\newcommand{\NwCl}{0.7}
\newcommand{\NwCf}{30.9}

\newcommand{\NwNewCf}{29.2}

\newcommand{\NcU}{0.77}
\newcommand{\NcP}{3.8}

\newcommand{\NcCp}{3.9}

\newcommand{\NcNewU}{0.8}

\newcommand{\NcMemo}{149.3}
\newcommand{\NcGap}{3.28}
\newcommand{\NcSens}{0.03}

\newcommand{\MaCp}{2.3}

\newcommand{\MaCf}{6.9}
\newcommand{\MaNewCp}{5.2}

\newcommand{\MaNewCf}{12.2}

\newcommand{\MbCp}{1.9}

\newcommand{\MbNewCp}{4.7}

\newcommand{\MbNewCf}{11.6}

\newcommand{\McCp}{2.3}

\newcommand{\McCf}{8.7}

\newcommand{\TrU}{0.24}
\newcommand{\TrP}{1.6}

\newcommand{\TrNewCp}{6.3}

\newcommand{\TrGf}{40.9}

\newcommand{\DonMs}{6.2}

\newcommand{\UnCp}{2.8}

\newcommand{\UnCf}{2.2}
\newcommand{\UnNewCp}{5.9}

\newcommand{\UnNewCf}{11.7}

\newcommand{\UnGf}{10.3}

\newcommand{\RnGf}{68.5}

\newcommand{\MgNewCp}{9.8}

\newcommand{\PnNewCp}{8.9}

\newcommand{\MlpMs}{6.9}

\newcommand{\HoGf}{6.5}

\newcommand{\HoMs}{7.1}
\newcommand{\HoCpu}{0.21}
\newcommand{\HoGap}{2.56}
\newcommand{\HoSens}{0.21}
\newcommand{\HoCpMean}{2.9}
\newcommand{\HoCfMean}{6.1}

\newcommand{\HoCpPm}{0.2}
\newcommand{\HoCfPm}{0.1}

\newcommand{\XiCfMean}{18.1}

\newcommand{\XiCfPm}{1.2}

\newcommand{\LocCpMean}{2.4}
\newcommand{\LocCfMean}{8.0}

\newcommand{\LocCpPm}{0.3}
\newcommand{\LocCfPm}{0.5}

\newcommand{\HoBandUA}{21.5}

\newcommand{\HoBandUC}{2.0}

\newcommand{\HoBandPE}{2.1}

\newcommand{\NwBandUA}{26.9}

\newcommand{\NwBandUC}{1.5}

\newcommand{\LocBandPE}{8.2}

\newcommand{\UnBandUA}{5.9}

\newcommand{\HoDistZero}{0.21}

\newcommand{\HoDistFour}{1.8}

\newcommand{\NcDistZero}{0.03}
\newcommand{\NcDistOne}{3.2}

\newcommand{\NcDistThree}{3.3}

\newcommand{\ShareThree}{88}

\newcommand{\ContRatio}{715}

\newcommand{\GlobLBOut}{24}

\newcommand{\GlobLDWall}{58}
\newcommand{\GlobLDOut}{42}
\newcommand{\WRatioMin}{122}
\newcommand{\WRatioMax}{2626}

\newcommand{\GamVarA}{0.07}
\newcommand{\GamVarU}{0.03}

\newcommand{\NfClTwenty}{0.58}
\newcommand{\SpecCpWall}{94.5}
\newcommand{\SpecUt}{91.1}

\newcommand{\RippleHigh}{2.3}

\newcommand{\SwWACp}{2.5}

\newcommand{\SwWACf}{9.4}

\newcommand{\SwMACp}{3.6}

\newcommand{\SwMAGf}{6.5}
\newcommand{\SwBaseCp}{2.5}

\newcommand{\SwBasePar}{12.7}

\newcommand{\SwMCCp}{2.3}

\newcommand{\SwMCPar}{28.5}
\newcommand{\SwMCGf}{6.6}
\newcommand{\SwWCCp}{2.5}

\newcommand{\SwWCCf}{7.6}

\newcommand{\GfLocalPct}{86}

\newcommand{\GfFftPct}{7}

\newcommand{\GfSpecPct}{1}

\newcommand{\TrainMin}{124}
\newcommand{\HwGpu}{RTX 3050 Laptop GPU}
\newcommand{\HwCpu}{AMD Ryzen 7 4800H}
\newcommand{\GapCorr}{0.92}

\newcommand{\GapCpA}{7.5}
\newcommand{\GapCpB}{3.7}
\newcommand{\GapCpC}{2.0}

\newcommand{\TrTrainMin}{571}

\newcommand{\HoTrainMin}{124}
\newcommand{\MgTrainMin}{1490}

\newcommand{\EtaCp}{0.33}
\newcommand{\EtaUt}{0.07}

\newcommand{\WfRiseA}{38}
\newcommand{\WfRiseD}{10}
\newcommand{\WfShareA}{3}
\newcommand{\WfShareB}{5}
\newcommand{\WfShareD}{10}
\newcommand{\WfScaledVar}{2.5}

\newcommand{\RfXiDirCp}{5.11}

\newcommand{\RfEtaDirCp}{3.23}

\newcommand{\RfBothDirCp}{5.28}

\newcommand{\RfBothDilCp}{2.70}
\newcommand{\RfBothDilU}{1.06}
\newcommand{\RfBothIntCp}{2.69}
\newcommand{\RfBothIntU}{0.93}
\newcommand{\RfTrainCp}{2.70}

\newcommand{\RfFloorU}{0.80}

\newcommand{\RfBothUnCp}{15.5}

\newcommand{\RfBothMgCp}{16.3}

\newcommand{\HoSepR}{0.92}
\newcommand{\HoNewSepR}{0.89}

\newcommand{\NwSepR}{$-$11.3}
\newcommand{\NwNewSepR}{$-$49.6}

\newcommand{\DoNewSepR}{0.83}

\newcommand{\MgNewSepR}{0.95}

\newcommand{\KnNewSepR}{0.85}

\newcommand{\HoSepRMean}{0.86}
\newcommand{\HoSepRPm}{0.06}
\newcommand{\XiSepRMean}{$-$1.25}
\newcommand{\XiSepRPm}{0.37}
\newcommand{\LocSepRMean}{$-$3.38}
\newcommand{\LocSepRPm}{0.64}
\newcommand{\BbSepRLo}{0.36}

\newcommand{\SwBaseSepR}{0.22}
\newcommand{\SwMASepR}{$-$0.28}

\newcommand{\SwDASepR}{$-$2.11}
\newcommand{\SwDCSepR}{0.64}
\newcommand{\SepLocUn}{0.008}
\newcommand{\AngSepRLo}{0.71}
\newcommand{\AngSepRHi}{0.92}
\newcommand{\NewClR}{0.986}

\begin{document}

\title{Inductive and learning biases of a geometry- and physics-encoded Fourier neural operator for turbulent airfoil flow prediction from sparse data}

\author{Ada Sinha}
\thanks{These authors contributed equally to this work.}
\affiliation{Department of Mechanical Engineering, Indian Institute of Technology Gandhinagar, Gandhinagar, Gujarat 382055, India}
\author{Kushagreek Basu}
\thanks{These authors contributed equally to this work.}
\affiliation{Department of Mechanical Engineering, Indian Institute of Technology Gandhinagar, Gandhinagar, Gujarat 382055, India}
\author{Shubhangi Bansude}
\email{shubhangi.bansude@iitgn.ac.in}
\affiliation{Department of Mechanical Engineering, Indian Institute of Technology Gandhinagar, Gandhinagar, Gujarat 382055, India}

\date{\today}

\begin{abstract}
Neural-operator surrogates can predict turbulent flow fields at a fraction of the cost of computational fluid dynamics, but the near-wall quantities that govern aerodynamic performance, the wall shear stress and the separation point, remain difficult to learn, particularly from sparse data. It is also unclear which prior knowledge, built into the network structure (inductive biases) or into the training loss (learning biases), makes a prediction accurate. We address this question with a geometry- and physics-encoded Fourier neural operator (\ours{}) that acts on a body-fitted grid and is trained on Reynolds-averaged Navier--Stokes solutions for different airfoils, each observed at only 5\% of its grid nodes. Without observing the test cases, \ours{} predicts the surface pressure, lift and skin friction at held-out flow conditions with errors of \HoCp\%, \HoCl\% and \HoCf\%, although the skin friction is never supervised. On eight unseen airfoils, its errors
are comparable to the discrepancy between the training solver and an independent one. Removing each bias in turn shows that a wall factor enforcing the no-slip condition determines whether skin friction and separation can be learned, that a continuity residual, the only constraint at unobserved nodes, reduces memorization of the sensor data more than tenfold, and
that two-dimensional spectral layers carry the global pressure coupling. Compared with seven other architectures trained identically, \ours{} requires the least computation per prediction, \HoGf{} billion floating-point operations, with equal or better surface-pressure accuracy. Accurate, inexpensive learning from sparse data thus stems from biases matched to the flow physics rather than from larger networks.
\end{abstract}

\maketitle

%=====================================================================
\section{Introduction}\label{sec:intro}
%=====================================================================

The aerodynamic performance of an airfoil is governed by the flow in the thin boundary layer next to its surface. Lift
follows from the surface-pressure distribution, whereas drag and stall depend on the skin friction and on the point at which
the boundary layer separates from the wall. At the chord Reynolds numbers of practical interest, $10^6$--$10^7$, this
boundary layer is turbulent, and its prediction rests on turbulence modeling. Scale-resolving simulations remain too
expensive for routine use at these Reynolds numbers, so the Reynolds-averaged Navier--Stokes (RANS) equations, closed by an
eddy-viscosity model such as the $k$-$\omega$ shear-stress-transport (SST) model,\cite{menter1994,pope2000} are the standard
tool of aerodynamic analysis. Even a RANS solution, however, requires a mesh that resolves the viscous sublayer at the wall,
and a single two-dimensional solution costs of the order of a core-hour. A lift polar or a shape-optimization loop requires
hundreds of such solutions,\cite{martins2022} which motivates surrogate models that return the flow field at a fraction of
this cost.

Machine learning has become an established tool in fluid mechanics,\cite{kutz2017,brunton2020,vinuesa2022} from
data-driven turbulence closures\cite{ling2016,duraisamy2019} to surrogates that return a flow field in a fraction of a
second. For airfoils, convolutional networks trained on image-like representations of the geometry predict RANS fields
around shapes they were not trained on,\cite{guo2016,bhatnagar2019,sekar2019,thuerey2020} and point-cloud, graph and
attention-based networks do so on unstructured nodes.\cite{kashefi2021,pfaff2021,wu2024transolver} Physics-informed neural
networks (PINNs) embed the governing equations in the loss instead;\cite{raissi2019,cai2021} they have been used to solve the
RANS equations from boundary data alone\cite{eivazi2022} and to predict airfoil flow fields at reduced
cost.\cite{compress2025} Public benchmarks such as AirfRANS\cite{bonnet2022airfrans} have made these approaches comparable,
and the retrospective of the 2024 competition on machine learning for computational fluid dynamics (CFD), held at the
Conference on Neural Information Processing Systems, identifies generalization to new airfoil shapes as a central
difficulty.\cite{yagoubi2025ml4cfd} Such comparisons are informative only when every model is set up, tuned and reported
with equal care,\cite{lu2022fair,mcgreivy2024} a principle we follow in Sec.~\ref{sec:compare}.

Operator learning casts the solver itself as the object to be learned: a map between function spaces, from the geometry
and the flow parameters to the flow field, so that a new case is predicted without
retraining.\cite{kovachki2023,lu2021deeponet,azizzadenesheli2024} The Fourier neural operator (FNO)\cite{li2021fno}
implements this map with layers that convolve over the whole domain, the convolution being evaluated as a product in
Fourier space at a cost of $\mathcal O(N\log N)$ for $N$ nodes. Such global convolutions resemble the Green's-function
integrals that solve elliptic problems, among them the Poisson equation that couples the pressure of an incompressible flow
to its velocity everywhere. FNOs have since been applied widely in fluid mechanics. For turbulence, they serve as surrogates
for large-eddy simulation of three-dimensional turbulence,\cite{li2022fnoles} and attention,\cite{peng2022attn,peng2023}
implicit recurrent layers and U-Net blocks\cite{li2023iufno} have been added to make long predictions accurate and stable;
U-shaped variants learn multiphase flow in porous media from a fraction of the data that convolutional networks
need.\cite{wen2022ufno} For airfoils, FNOs predict steady flow fields,\cite{dai2022fournet} the unsteady flow of dynamic
stall\cite{meng2023stall} and, as Koopman neural operators, transonic buffet;\cite{meng2024buffet} learned deformations
extend them to general geometries,\cite{li2023geofno} and physics-informed losses reduce the data they
need.\cite{li2024pino}

Three properties of the FNO matter for wall-bounded flows. First, the Fourier transform presupposes a periodic rectangle of
uniformly spaced points, so a flow around a body must be mapped onto such a domain, through learned
deformations,\cite{li2023geofno} encoders from an irregular mesh to a latent grid,\cite{li2023gino} or body-fitted
coordinates as in physics-informed convolutional networks.\cite{gao2021phygeonet} Second, truncating the Fourier series
biases the operator toward low wavenumbers, in line with the spectral bias of neural networks,\cite{rahaman2019,qin2024}
whereas the steepest gradients of an airfoil flow lie in the thin layer next to the wall. Third, the independence of the
discretization often attributed to FNOs holds for the operator in function space but only approximately for its discrete
implementation, in which aliasing and local layers break it.\cite{bartolucci2023,raonic2023} How an FNO behaves next to a
wall, under each of these constraints, is therefore not settled by its success on periodic or far-field problems.

Practical data add two further difficulties. The first concerns the data themselves. A CFD solution is dense, but training
on every node of every solution of a large simulation database is costly in memory and time, and such databases often
store only part of each field. Experimental data are sparse by nature: a wind-tunnel model carries a few tens of pressure
taps, and particle image velocimetry covers a window of the flow. Models that learn from sparse observations are therefore
valuable beyond the saving in training cost, because they can be trained on multimodal data that combine dense simulations
with sparse measurements, the combination on which digital twins of engineering systems are being built\cite{rasheed2020}
and which multi-fidelity learning exploits.\cite{meng2020composite} Sparse-reconstruction methods complete the field of one
flow from point observations of that same flow, through sparse representations,\cite{callaham2019} shallow or convolutional
networks,\cite{erichson2020,fukami2021,jin2018} graph networks\cite{duthe2023} or physics-informed
networks.\cite{raissi2020hfm,hosseini2024} These methods are typically formulated to reconstruct a configuration from its own
observations, rather than to predict configurations that were never observed. Our setting is different: every solution of a
database is observed at a fixed 5\% of its nodes, and new conditions and new airfoils are then predicted from the geometry
and the flow parameters alone. The 95\% of each training field that lies between the sensors must then be constrained by
something other than data. Physics-informed learning supplies such a constraint by penalizing equation residuals at
unobserved points,\cite{raissi2019,sun2020,wang2021pideeponet,li2024pino} but soft penalties are hard to balance against the
data and to optimize, and they guarantee nothing;\cite{wang2021gradient,krishnapriyan2021} hard constraints that build a
boundary condition into the form of the output avoid both problems for the conditions they cover.\cite{lagaris1998,lu2021hpinn,sukumar2022,saad2023boon}

The second difficulty is the wall. The pressure coefficient is a \emph{value} of the field at the wall, whereas the wall
shear stress is a \emph{derivative} taken across a viscous sublayer about $10^{-4}c$ thick, where $c$ is the airfoil chord,
and it is correspondingly sensitive to errors in the near-wall velocity. On AirfRANS, surrogates that predict the fields
well struggle to infer the wall shear stress, and their drag is unreliable;\cite{bonnet2022airfrans} skin friction and
separation are rarely learned or tested. Machine learning for wall-bounded turbulence has concentrated instead on wall
models for large-eddy simulation\cite{yang2019} and on estimating the flow above the wall from measurements at the
wall,\cite{guastoni2021,guemes2021} the opposite direction of inference.

Prior knowledge enters a learned model of a physical system in three ways.\cite{karniadakis2021} Observational biases
are carried by the data. Inductive biases are built into the structure of the model, so that a property holds for any value
of its trainable parameters. Learning biases are imposed through the loss, so that a property is favored during training but
not guaranteed. Operators for fluid flow usually combine all three, yet it has rarely been tested which of them an FNO needs
when most of every training field is unobserved, or how each one affects the predicted flow. We address this question by
building an FNO for this setting with encodings of the geometry and of the physics: a body-fitted grid that unrolls the
domain into a rectangle, an output factor that builds no-slip into the velocity, and residuals of conservation laws on the
unobserved nodes. The resulting operator, which we call \ours{} (geometry- and physics-encoded FNO), learns solutions of the
$k$-$\omega$ SST model for three airfoils of the four-digit series of the National Advisory Committee for Aeronautics
(NACA) from 5\% of each training field and derives its skin friction and separation point from the predicted field without
ever being trained on them. Its observational bias is the fixed 5\% of each field; its inductive biases are the body-fitted
grid, the global and local paths of its layers, the truncated spectrum, the conditioning on the flow parameters through
feature-wise linear modulation (FiLM)\cite{perez2018film} and the wall factor; its learning biases are the residuals of
conservation laws. Each bias is then removed or altered in turn under one training protocol, its effect is quantified in the
predicted flow fields, their error fields and the wall quantities, and its mechanism is traced inside the trained network.
Finally, the Fourier backbone is replaced by seven other architectures that receive the same inputs and training, and the
operators are compared in accuracy and in the cost of one prediction.

The contributions are as follows.
(i) A Fourier neural operator that learns from partial fields and predicts the surface pressure, lift, skin friction and
separation of new airfoils, tested against an independent RANS solver whose own disagreement with the training data is
measured separately.
(ii) An investigation of its inductive biases, the two-dimensional global path, the retained modes and the capacity, the
local path and the transfer to finer grids, the parameter conditioning and the wall factor, and of its learning biases, the
continuity and momentum residuals, each with its effect on the flow and its mechanism.
(iii) A comparison with seven other architectures, in which \ours{} needs the least arithmetic per prediction while its
surface pressure is as accurate as the best on the held-out conditions and the most accurate on new airfoils.
(iv) A roadmap that ties each bias to the property of the flow or of the numerics that motivates it and to its measured
effect, for the design of sparse-data operators of wall-bounded flows.

Section~\ref{sec:problem} describes the methodology: the flow problem and the data, the operator and its biases, the
evaluation protocol and the comparison architectures. Section~\ref{sec:results} presents and discusses the results: the
predictions of the complete operator, the role of each inductive and learning bias, a synthesis of their effects and the
comparison with other architectures. Section~\ref{sec:conclusions} summarizes the conclusions.

%=====================================================================
\section{Methodology}\label{sec:problem}
%=====================================================================

This section describes the flow problem and its quantities of interest (Sec.~\ref{sec:equations}), the training database
(Sec.~\ref{sec:database}), the independent solver and the unseen airfoils used for testing (Sec.~\ref{sec:reference}) and
the sparse observations (Sec.~\ref{sec:sparse}). It then presents the operator and its biases (Sec.~\ref{sec:operator}), the
evaluation protocol (Sec.~\ref{sec:protocol}) and the architectures against which the operator is compared
(Sec.~\ref{sec:baselines}).

\subsection{Governing equations and quantities of interest}\label{sec:equations}

We consider steady, incompressible, two-dimensional flow of air around an airfoil of chord $c$, at angle of attack
$\alpha$ and free-stream speed $\Uinf$. The Reynolds-averaged velocity $\mathbf u=(u,v)$ and pressure $p$ satisfy
\begin{equation}
\begin{gathered}
\nabla\cdot\mathbf u=0,\\
(\mathbf u\cdot\nabla)\mathbf u=-\frac{1}{\rho}\nabla p+\nabla\cdot\bigl[\nueff\bigl(\nabla\mathbf u+\nabla\mathbf u^{\mathsf T}\bigr)\bigr],
\end{gathered}
\label{eq:rans}
\end{equation}
where $\nueff=\nu+\nu_t$ and the eddy viscosity $\nu_t$ is supplied by the $k$-$\omega$ shear-stress-transport (SST)
model.\cite{menter1994} The flow obeys no-slip at the wall and free-stream conditions far from it, and for one shape it
is governed by $\alpha$ and the chord Reynolds number $Re=\Uinf c/\nu$.

The quantities of aerodynamic interest are evaluated at the wall. With $\mathbf n$ the outward unit normal, $\mathbf t$ the
unit tangent and $y_n$ the wall-normal distance,
\begin{equation}
\begin{gathered}
C_p=\frac{p-p_\infty}{\tfrac12\rho\Uinf^2},\qquad
C_L=-\frac{1}{c}\oint C_p\,(\mathbf n\cdot\mathbf e_L)\,\mathrm ds,\\
\tau_w=\mu\left.\frac{\partial(\mathbf u\cdot\mathbf t)}{\partial y_n}\right|_{y_n=0},\qquad
C_f=\frac{\tau_w}{\tfrac12\rho\Uinf^2},
\end{gathered}
\label{eq:wallq}
\end{equation}
with $\mathbf e_L=(-\sin\alpha,\cos\alpha)$; the small viscous contribution to lift is neglected. The boundary layer
separates where the suction-side wall shear changes sign. The pressure coefficient is a \emph{value} of the field at the
wall, but the wall shear is a \emph{derivative} taken across a thin viscous layer next to the wall, about $10^{-4}c$ thick; it is the hardest
quantity to learn from sparse data.

\subsection{Training database}\label{sec:database}

\begin{figure*}[!tbp]
\centering
\includegraphics[width=\textwidth]{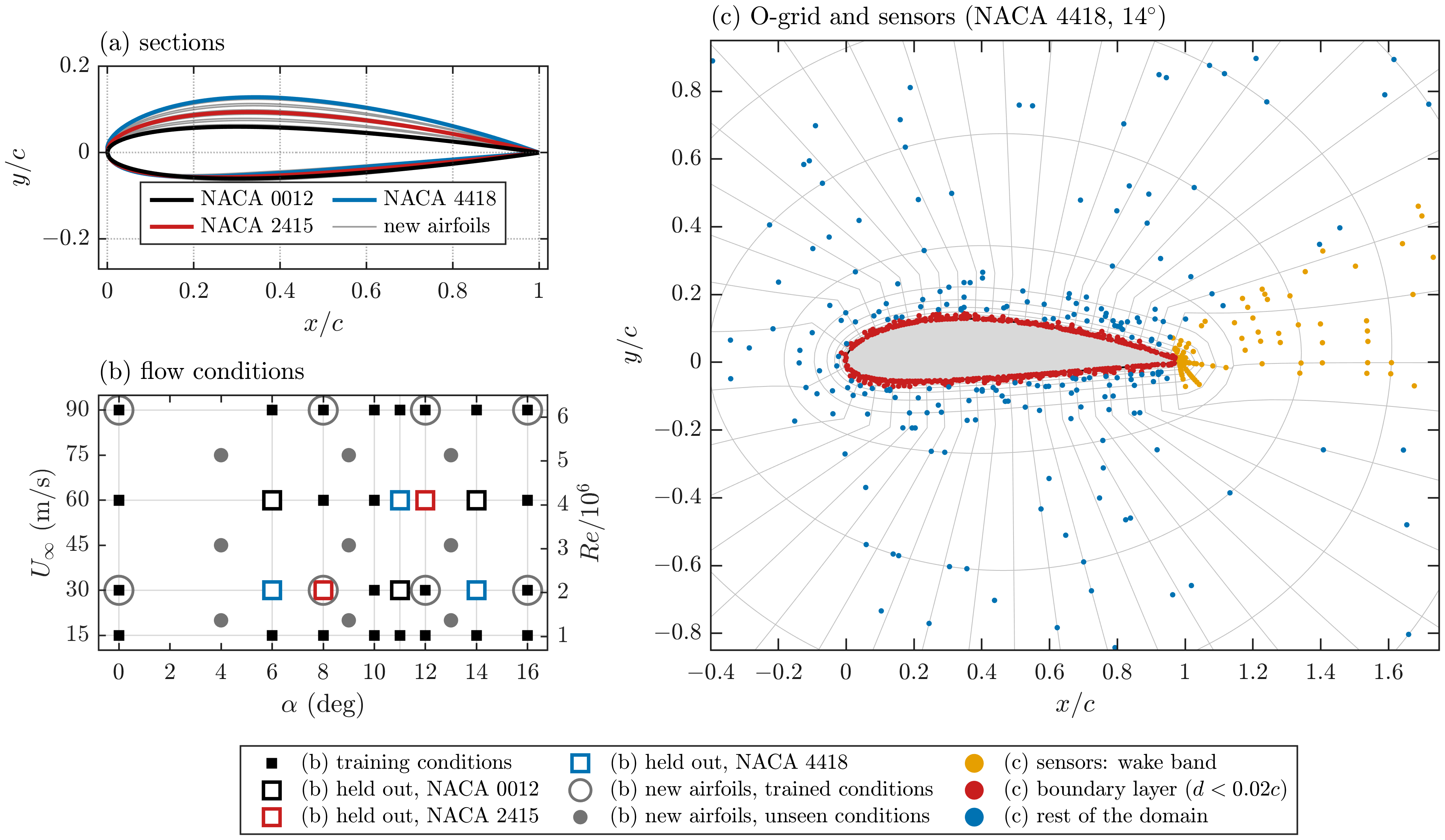}
\caption{\label{fig:geometry}Geometry, conditions and observations. (a) Sections of the three training airfoils and the
eight new airfoils. (b) Flow conditions in the plane of angle of attack and free-stream speed (right axis: chord Reynolds
number): the training lattice of eight angles and four speeds (gray lines); the held-out conditions, each withheld for the
airfoil of its color only; the trained conditions at which the new airfoils are solved (rings), where only the shape is new;
and the unseen conditions (dots), which lie between the lattice lines, so that shape, angle and speed are all new. (c) The
body-fitted O-grid of NACA~4418 (every second to fourth row near the wall and every sixth column drawn) and the 1024 fixed
sensors of the held-out condition at $14^\circ$, 30\,m/s, by stratum.}
\end{figure*}

The training database comprises 96 steady RANS solutions of the NACA 0012, 2415 and 4418 airfoils, each at eight angles
between $0^\circ$ and $16^\circ$ and four speeds between 15 and 90\,m/s (Table~\ref{tab:conditions}). The chord is
$c=1$\,m and the fluid is standard air, so that $Re=1.0$--$6.2\times10^6$ [Fig.~\ref{fig:geometry}(a, b)].

\begin{figure*}[!tbp]
\centering
\includegraphics[width=\textwidth]{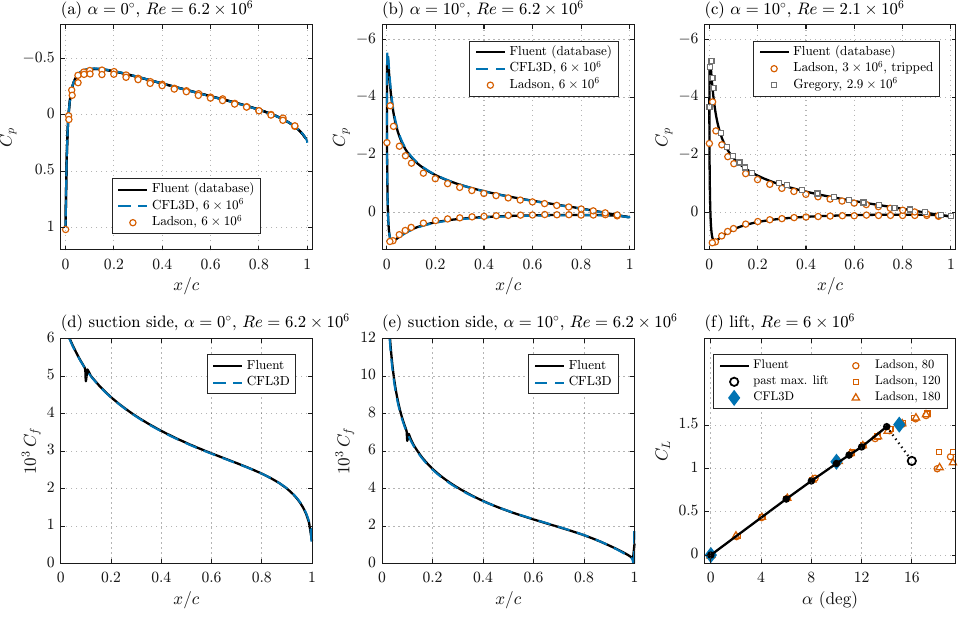}
\caption{\label{fig:validation}Validation of the training code: the Fluent solutions of NACA~0012 in the database against
the NASA CFL3D solution with the same turbulence model and against experiment. (a, b) Surface pressure at 90\,m/s
($Re=6.2\times10^6$), $\alpha=0^\circ$ and $10^\circ$, with CFL3D and Ladson's measurements at $6\times10^6$. (c) Surface
pressure at 30\,m/s ($Re=2.1\times10^6$), $\alpha=10^\circ$, with Ladson's tripped measurements at $3\times10^6$ and those
of Gregory and O'Reilly at $2.9\times10^6$. (d, e) Suction-side skin friction at $6.2\times10^6$, with CFL3D. (f) Lift at
$6.2\times10^6$, with CFL3D and Ladson's measurements at $6\times10^6$ for three sizes of trip grit (80, 120 and 180); open circle: the
Fluent solution past maximum lift, which does not settle to a steady state.}
\end{figure*}

The solutions were computed with ANSYS Fluent ($k$-$\omega$ SST, pressure-based coupled solver, second-order
discretization) on structured meshes of $2.0$--$4.9\times10^5$ cells (Appendix~\ref{app:data}). Before the database is
used as the reference for training and evaluation, its NACA~0012 solutions are compared with experiment and with the solution of the same turbulence
model by CFL3D, the structured-grid flow solver of the National Aeronautics and Space Administration (NASA),\cite{rumsey2010} at the Reynolds number of the measurements (Fig.~\ref{fig:validation}). At 90\,m/s,
$Re=6.2\times10^6$, the surface pressure follows CFL3D to a median difference of 0.002 in $C_p$ at $0^\circ$ and 0.012 at
$10^\circ$, and both follow the measurements of Ladson,\cite{ladson1988} including the suction peak
[Fig.~\ref{fig:validation}(a, b)]. The agreement holds at the low end of the database as well: at 30\,m/s and $10^\circ$
the pressure follows the measurements of Gregory and O'Reilly\cite{gregory1970} and of Ladson at $3\times10^6$
[Fig.~\ref{fig:validation}(c)]. The skin friction, the quantity that is hardest to learn, matches the CFL3D distribution
along the suction side to a median of 0.2\% at $0^\circ$ and 0.7\% at $10^\circ$, and to within 4.4\% everywhere between
$x/c=0.05$ and 0.95 [Fig.~\ref{fig:validation}(d, e)]. The lift is within 0.3\% of Ladson's tripped measurements for
$\alpha=6$--$12^\circ$ and 4.5\% above them at $14^\circ$ [Fig.~\ref{fig:validation}(f)]. Beyond $14^\circ$ the Fluent
solution stalls, about three degrees earlier than the experiment, which reaches its maximum lift near $17^\circ$ (open
circle). Past maximum lift the separated flow is unsteady, and a steady RANS solution is least reliable there. The solution at
$16^\circ$ does not converge to a steady state, and its integral momentum balance closes only to 2.5\%, compared with 0.5\%
at the angles below $14^\circ$; here the balance is measured as the difference between the drag integrated over the wall and
over a contour far from it. The few database solutions past maximum lift (Appendix~\ref{app:data}) therefore represent the
steady turbulence model rather than the measured stall, and Sec.~\ref{sec:res_new} shows that the operator inherits this
early stall.

\begin{table*}[!tp]
\caption{\label{tab:conditions}Flow conditions. Held-out conditions are given as $(\alpha,\Uinf)$ and are withheld for the
airfoil named only. New airfoils: NACA 0013, 1413, 1414, 2414, 2416, 3416, 3417 and 4417. With $c=1$\,m,
$Re/10^6=1.0$, 1.4, 2.1, 3.1, 4.1, 5.1 and 6.2 at $\Uinf=15$, 20, 30, 45, 60, 75 and 90\,m/s.}
\begin{ruledtabular}
\begin{tabular}{lllllr}
Set & Airfoils & Code & $\alpha$ (deg) & $\Uinf$ (m/s) & Solutions \\
\hline
Training database & NACA 0012, 2415, 4418 & Fluent & 0, 6, 8, 10, 11, 12, 14, 16 & 15, 30, 60, 90 & 96 \\
\quad held out & NACA 0012 & Fluent, OpenFOAM & \multicolumn{2}{l}{$(6^\circ, 60)$, $(11^\circ, 30)$, $(14^\circ, 60)$} & 3 \\
 & NACA 2415 & Fluent, OpenFOAM & \multicolumn{2}{l}{$(8^\circ, 30)$, $(12^\circ, 60)$} & 2 \\
 & NACA 4418 & Fluent, OpenFOAM & \multicolumn{2}{l}{$(6^\circ, 30)$, $(11^\circ, 60)$, $(14^\circ, 30)$} & 3 \\
\quad training set & NACA 0012, 2415, 4418 & Fluent & \multicolumn{2}{l}{all others} & 88 \\[2pt]
New airfoils, trained conditions & eight new airfoils & OpenFOAM & 0, 8, 12, 16 & 30, 90 & \CondTrN \\
New airfoils, unseen conditions & eight new airfoils & OpenFOAM & 4, 9, 13 & 20, 45, 75 & \CondUnN \\
\end{tabular}
\end{ruledtabular}
\end{table*}

Every solution is interpolated onto the body-fitted grid of Sec.~\ref{sec:geometry}; we call the interpolated fields
$(u,v,p,\nueff)$ the \emph{reference field}. Surface quantities of the reference are computed from it with the same
discrete formulas as those of a prediction (Sec.~\ref{sec:supervision}), so that a reported error is a difference between
two fields.

\subsection{Independent solver and unseen airfoils}\label{sec:reference}

Generalization is tested most stringently against a reference that shares no systematic modeling or numerical choice with
the training data. Eight new airfoils, NACA 0013, 1413, 1414, 2414, 2416, 3416, 3417 and 4417 [Fig.~\ref{fig:geometry}(a)],
were therefore solved with a different code, OpenFOAM v2312\cite{weller1998openfoam} (\texttt{simpleFoam}, $k$-$\omega$ SST,
a $1.48\times10^5$-cell C-grid with $y^+\approx1$), at two sets of conditions [Fig.~\ref{fig:geometry}(b)]: at the \emph{trained conditions}
$(0,8,12,16)^\circ\times(30,90)$\,m/s, which belong to the training lattice, so that only the shape is new, and at the
\emph{unseen conditions} $(4,9,13)^\circ\times(20,45,75)$\,m/s, at which shape, angle and speed all differ from the
training data. This gives \Nnew{} solutions in total (Appendix~\ref{app:openfoam}).

\begin{figure*}[!tbp]
\centering
\includegraphics[width=\textwidth]{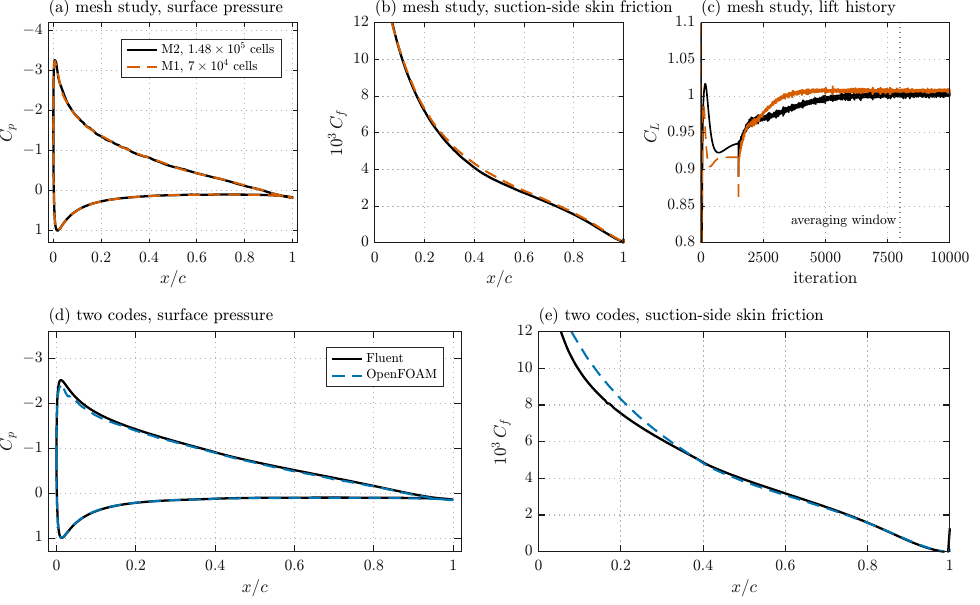}
\caption{\label{fig:codeval}Validation of the second code. (a--c) Mesh study of OpenFOAM on NACA~1414 at $9^\circ$,
45\,m/s ($Re=3.1\times10^6$): surface pressure, suction-side skin friction and lift history on the coarse (M1) and fine
(M2, used throughout) meshes; the dotted line marks the start of the averaging window. (d, e) Fluent and OpenFOAM on the
held-out condition of NACA~2415 at $8^\circ$, 30\,m/s.}
\end{figure*}

The second solver is validated before it is used as a reference (Fig.~\ref{fig:codeval}). On NACA~1414 at $9^\circ$ and
45\,m/s, refining the mesh from $7\times10^4$ to $1.48\times10^5$ cells changes the surface pressure by 1.8\% and the skin
friction by 1.9\% (relative norms), the lift by 0.5\% and the drag by 4.6\%; the two pressure and friction distributions
are nearly indistinguishable, and the lift of both meshes has settled before the last 2000 iterations, over which the
fields are averaged [Fig.~\ref{fig:codeval}(a--c)]. The eight held-out conditions of the training airfoils were then solved
with OpenFOAM as well. On NACA~2415 at $8^\circ$ and 30\,m/s the two codes give the same pressure distribution apart from a
slightly lower suction peak in OpenFOAM, and the same skin friction aft of $x/c\approx0.35$, ahead of which OpenFOAM's is
higher [Fig.~\ref{fig:codeval}(d, e)]. Over the eight conditions the two codes, with the same turbulence model, agree to a
median of \GapU\% in $u$, \GapCp\% in $C_p$ and \GapCl\% in $C_L$; the lift differs by less than 5\% at seven of them and
by 7.1\% at the eighth, NACA~0012 at $14^\circ$ and 60\,m/s, close to stall. We call this residual difference, which reaches \GapP\% in $p$ and \GapCf\% in $C_f$, the \emph{code gap}. It involves
no learned model, and it sets the scale against which the errors of an operator trained on Fluent solutions and evaluated
against OpenFOAM should be compared (Sec.~\ref{sec:res_new}).

\subsection{Sparse observations}\label{sec:sparse}

Each training solution is observed at a fixed, never-resampled set of 1024 nodes, 5\% of the 20\,480 nodes of the grid
[Fig.~\ref{fig:geometry}(c)]. The draw is stratified by geometry and angle alone, never by the flow: a wake band that leaves
the trailing edge in the free-stream direction is sampled at 25\%, nodes within $0.02c$ of the wall at 5\%, and the
remaining domain takes the rest of the budget. The boundary layer thus receives about half of the sensors, only about two
per grid column. A solution carries $4\times20\,480=81\,920$ values, of which 4096 are observed; the remaining 95\% of every training field
is constrained only by the model and its biases. When a new case is predicted, no observation of it is used.

%=====================================================================
\subsection{Geometry- and physics-encoded Fourier neural operator}\label{sec:operator}
%=====================================================================

This subsection describes \ours{} and its biases. Each bias is motivated by a specific property of the flow or of the
numerics, and its contribution to the trained operator is quantified in Secs.~\ref{sec:inductive} and \ref{sec:learning}.
Table~\ref{tab:ingredients} lists the biases together with the variation of each that is tested.

\begin{figure*}[!tbp]
\centering
\includegraphics[width=\textwidth]{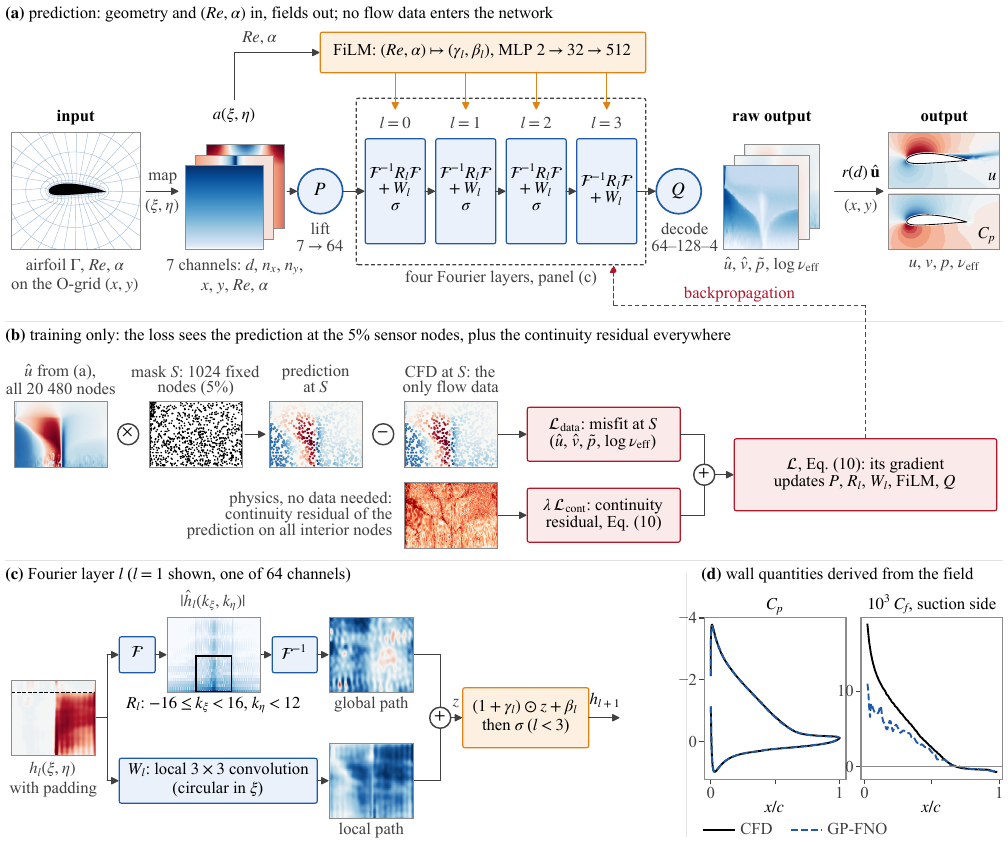}
\caption{\label{fig:arch}\ours{}, drawn with the data of the held-out NACA~4418 at $\alpha=14^\circ$, 30\,m/s. (a)
Prediction: seven geometry and parameter inputs on the computational plane (three shown), lifting $P$, four Fourier
layers modulated by feature-wise linear modulation (FiLM), decoder $Q$ and wall factor $r(d)$; no flow value is an input. (b) Training only: the
prediction is masked by the fixed sensor set $S$ (1024 nodes, 5\%) and compared with the reference there, and the
continuity residual is evaluated on all interior nodes. (c) One Fourier layer, Eq.~(\ref{eq:layer}), with one feature
channel of the trained model; box: retained modes; dashed line: the 16 padded rows. (d) $C_p$ and suction-side $C_f$
derived from the predicted field.}
\end{figure*}

Figure~\ref{fig:arch} summarizes the operator. In prediction [Fig.~\ref{fig:arch}(a)], seven inputs that describe the
geometry and the flow parameters are defined on the body-fitted O-grid and mapped to the rectangular computational plane
$(\xi,\eta)$. A pointwise layer $P$ lifts them to 64 channels, four Fourier layers modulated by FiLM process them, and a
pointwise decoder $Q$ returns four raw outputs, of which the velocity is multiplied by the wall factor $r(d)$ to give the
physical velocity; no flow quantity is an input. In training [Fig.~\ref{fig:arch}(b)], the prediction is compared with the
reference only at the 1024 sensor nodes of each case, the continuity residual is evaluated at all interior nodes, and the
gradient of the combined loss updates all trainable parameters. Figure~\ref{fig:arch}(c) details one Fourier layer, with
its global (spectral) and local (convolutional) paths, and Fig.~\ref{fig:arch}(d) shows the surface pressure and skin
friction derived from the predicted field.

Sections~\ref{sec:geometry}--\ref{sec:wall_method} concern \emph{inductive biases}. An inductive bias is an assumption
about the solution that is built into the structure of the model, into the grid on which it acts, the operations of its
layers or the form of its output, so that it holds for any value of the trainable parameters and for every input,
independently of the training data.\cite{karniadakis2021} A familiar example is the convolutional layer: it applies the same
kernel at every location and therefore assumes translation invariance, i.e., that the same local pattern can occur anywhere
in the domain. Another example is a network output multiplied by a function that vanishes on a boundary, which enforces a
homogeneous Dirichlet condition exactly. An inductive bias narrows the set of functions the model can represent; it helps
when the solution lies in the narrowed set and harms when it does not. The learning biases are described in
Sec.~\ref{sec:supervision}.

\begin{table*}[!tbp]
\caption{\label{tab:ingredients}The inductive and learning biases of \ours{}, the idea behind each, and the change made to it in the
investigation. Every changed operator is retrained from scratch under the protocol of Sec.~\ref{sec:protocol}.}
\renewcommand{\arraystretch}{1.1}
\begin{ruledtabular}
\begin{tabular}{llll}
Bias & Motivation & In \ours{} & Variation tested (section) \\
\hline
\multicolumn{4}{l}{\emph{Inductive biases}}\\
\quad Computational plane & \parbox[t]{2.1in}{\raggedright put the wall on one grid row and the resolution where the gradients are\strut} & \parbox[t]{1.55in}{\raggedright $256\times80$ O-grid, $d_1=10^{-5}c$\strut} & \parbox[t]{1.6in}{\raggedright shared by every model; not removed\strut} \\
\quad Global path & \parbox[t]{2.1in}{\raggedright pressure at a point depends on the flow everywhere\strut} & \parbox[t]{1.55in}{\raggedright two-dimensional Fourier layers\strut} & \parbox[t]{1.6in}{\raggedright along $\xi$ only; removed (\ref{sec:spectral})\strut} \\
\quad Local path & \parbox[t]{2.1in}{\raggedright features finer than the retained modes\strut} & \parbox[t]{1.55in}{\raggedright $3\times3$ convolutions\strut} & \parbox[t]{1.6in}{\raggedright refined grids, kernels dilated (\ref{sec:resolution})\strut} \\
\quad Retained modes & \parbox[t]{2.1in}{\raggedright keep the scales that the wall spectrum needs\strut} & \parbox[t]{1.55in}{\raggedright $-16\le k_\xi<16$, $0\le k_\eta<12$\strut} & \parbox[t]{1.6in}{\raggedright halved, raised by half (\ref{sec:capacity})\strut} \\
\quad Width and depth & \parbox[t]{2.1in}{\raggedright capacity of the operator\strut} & \parbox[t]{1.55in}{\raggedright 64 channels, four layers\strut} & \parbox[t]{1.6in}{\raggedright 32 and 128 channels; three and six layers (\ref{sec:capacity})\strut} \\
\quad Conditioning & \parbox[t]{2.1in}{\raggedright $Re$ and $\alpha$ act on the whole flow at once\strut} & \parbox[t]{1.55in}{\raggedright FiLM in every layer\strut} & \parbox[t]{1.6in}{\raggedright removed (\ref{sec:film})\strut} \\
\quad Wall factor & \parbox[t]{2.1in}{\raggedright no-slip holds exactly, and the wall shear becomes a value of the output\strut} & \parbox[t]{1.55in}{\raggedright $\mathbf u=r(d)\,\hat{\mathbf u}$\strut} & \parbox[t]{1.6in}{\raggedright removed (\ref{sec:wall})\strut} \\
\multicolumn{4}{l}{\emph{Learning biases}}\\
\quad Continuity residual & \parbox[t]{2.1in}{\raggedright mass is conserved at nodes that carry no data\strut} & \parbox[t]{1.55in}{\raggedright relative residual, $\lambda=0.1$\strut} & \parbox[t]{1.6in}{\raggedright removed (\ref{sec:continuity})\strut} \\
\quad Momentum residual & \parbox[t]{2.1in}{\raggedright the momentum balance also holds between the sensors\strut} & \parbox[t]{1.55in}{\raggedright not used\strut} & \parbox[t]{1.6in}{\raggedright added at $w=0.01$, 0.03, 0.1 (\ref{sec:momentum})\strut} \\
\end{tabular}
\end{ruledtabular}
\end{table*}

\subsubsection{Body-fitted computational plane}\label{sec:geometry}

The Fourier layer at the core of an FNO represents each feature field by a truncated Fourier series, which requires a
rectangular array of equally spaced points that is periodic in each direction. The domain around an airfoil is neither rectangular nor periodic, and its
wall, where no-slip holds and the wall shear is defined, is curved. As in the geometry-aware FNO,\cite{li2023geofno} we
therefore map the domain onto a rectangle, here with a body-fitted O-grid of $256\times80$ nodes.\cite{thompson1985} One index
runs around the airfoil and the other outward from it, so that the wall becomes the bottom row of the rectangle and the outer
boundary, a circle of radius $2c$, its top row [Fig.~\ref{fig:geometry}(c)]. Around the airfoil the grid closes on itself,
so every field is periodic in that direction, as the Fourier transform assumes. In the wall-normal direction the field is not periodic; the
network therefore pads it with 16 rows of zeros beyond the outer boundary [dashed line in Fig.~\ref{fig:arch}(c)], so that
the discrete transform does not couple the wall row to the outer boundary. The rows are clustered at the wall, where the gradients are: the
first lies at $d_1=10^{-5}c$, 44 of the 80 rows lie within $0.02c$ of the wall, and along the surface the spacing is $0.008c$.
Every airfoil, trained or new, is meshed by the same rule. Each node carries seven inputs, all known before any flow is
computed: the wall distance $d$ and the position $(x,y)$, both normalized by the outer radius $2c$, the wall normal
$(n_x,n_y)$, and the flow parameters $Re/10^6$ and $\alpha/10^\circ$. Derivatives in the physical plane, which the residuals and the wall shear need, follow from the grid by the
chain rule.

\subsubsection{Global and local paths}\label{sec:paths}
The global path is motivated by the elliptic nature of the pressure equation in incompressible flow: the pressure satisfies a Poisson equation, so its value at any point depends on the flow throughout the domain. Taking the
divergence of the momentum equation in Eq.~(\ref{eq:rans}) and using continuity yields a Poisson equation for the pressure,
\begin{equation}
\frac{1}{\rho}\nabla^2p=-\partial_iu_j\,\partial_ju_i+2\,\partial_i\partial_j\bigl(\nueff S_{ij}\bigr),
\label{eq:poisson}
\end{equation}
where $S_{ij}$ is the strain rate. The solution of a Poisson equation is an integral of its source over the whole domain,
weighted by a Green's function, so the pressure at a point depends on the flow everywhere. A network built only from local
stencils passes information one neighbor per layer and would need tens of layers to connect the wall with the outer flow.

A Fourier layer can be interpreted as a learned approximation of such an integral operator.\cite{kovachki2023} It transforms
each feature field into Fourier space, multiplies the retained Fourier modes by learned complex weights that also mix the
channels, and transforms the result back; because every Fourier mode spans the whole domain, a single layer couples every node to every other, at a cost of $\mathcal O(N\log N)$ for $N$ nodes.\cite{li2021fno}
\ours{} lifts its seven inputs to a 64-channel field $h_0$ with a pointwise layer $P$ and updates it in four layers,
$l=0,\dots,3$,
\begin{equation}
\begin{gathered}
h_{l+1}=\sigma\bigl((1+\gamma_l)\odot\bigl[G_l+L_l\bigr]+\beta_l\bigr),\\
G_l=\mathcal F^{-1}\bigl(R_l\cdot\mathcal Fh_l\bigr),\qquad L_l=W_lh_l.
\end{gathered}
\label{eq:layer}
\end{equation}
Here $\mathcal F$ is the two-dimensional discrete Fourier transform and $R_l$ a learned complex $64\times64$ matrix for each
retained mode, so that $G_l$ is the \emph{global path}. The \emph{local path} $L_l$ is a small convolution, $1\times1$ in the
first layer and $3\times3$ thereafter, much like a finite-difference stencil. The global path carries the coupling over
distance, and the local path the features finer than the retained modes. The channel-wise scale $\gamma_l$ and shift $\beta_l$ are computed
from the flow parameters by FiLM (Sec.~\ref{sec:film_method}), and $\sigma$ is the Gaussian error linear unit (GELU) activation,\cite{hendrycks2016gelu} omitted in the
last layer. A pointwise decoder $Q$ (64--128--4 channels) returns four raw outputs (Sec.~\ref{sec:wall_method}). The network
has $1.27\times10^7$ parameters, nearly all of them in the matrices $R_l$.

\subsubsection{Retained Fourier modes selected from the energy spectra}\label{sec:modes_method}

Any quantity $f_i$ sampled at $N$ equally spaced points around the airfoil is represented exactly by its discrete Fourier
series,
\begin{equation}
f_i=\sum_{k=-N/2}^{N/2-1}\hat f_k\,e^{2\pi\mathrm i\,ki/N},
\label{eq:fourier}
\end{equation}
where the wavenumber $k$ is the number of periods of the corresponding Fourier mode along one closed grid line around the
airfoil. Resolving a feature of width $w$ on a perimeter $P\approx2c$ requires modes up to $|k|\sim P/w$: about ten for the
pressure at half a chord from the wall, but fifty or more for a suction peak whose width is a few hundredths of the chord.
Retaining modes is costly, however: every retained mode costs $2\times64^2=8192$ parameters in each layer, and retaining all $256\times49$ modes of the padded field would cost
$4\times10^8$ parameters, whereas the $32\times12=384$ modes retained by \ours{} cost $3.1\times10^6$ per layer. Truncation
therefore makes the global path affordable, and the cutoff must be chosen according to the modes that the flow requires.

This choice can be made from the reference data before training. By Parseval's theorem,
\begin{equation}
E_k=\frac{|\hat f_k|^2+|\hat f_{-k}|^2}{\sum_{k'\neq0}|\hat f_{k'}|^2},\qquad
\frac{\|f-f_K\|}{\|f-\bar f\|}=\Bigl(\sum_{k>K}E_k\Bigr)^{1/2}
\label{eq:parseval}
\end{equation}
is the share of the variation carried by the pair $\pm k$, and the energy beyond a cutoff $K$ is exactly the squared relative
error of the best representation truncated there, $f_K$. The spectrum of the reference fields therefore states what a given
cutoff must miss.

\begin{figure}[!htbp]
\centering
\includegraphics[width=\columnwidth]{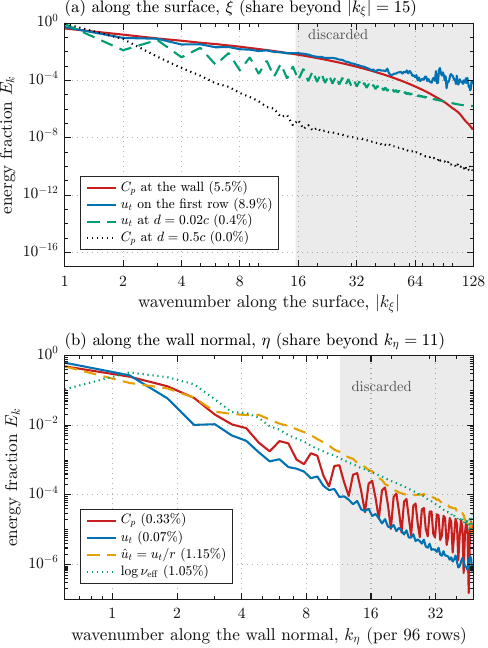}
\caption{\label{fig:spectra}Energy fraction $E_k$, Eq.~(\ref{eq:parseval}), of the reference fields, averaged over the
eight held-out conditions. (a) Along $\xi$: wall pressure, tangential velocity on the first row and at $d=0.02c$, and
pressure at $d=0.5c$. (b) Along $\eta$, over every grid column, from the cosine transform of the column: pressure,
physical and scaled tangential velocity (Sec.~\ref{sec:wall_method}) and effective viscosity. Shaded: wavenumbers beyond
the retained $|k_\xi|\le15$ and $k_\eta\le11$; the legends give the share of energy there.}
\end{figure}

Along the surface, the pressure at half a chord from the wall carries practically no energy beyond $|k_\xi|=10$
[Fig.~\ref{fig:spectra}(a)]. Closer to the wall the spectra become flatter, because the wall pressure contains the suction
peak and the first-row velocity contains the separation point, both features only a few hundredths of the chord long. Modes
up to $|k_\xi|=15$ contain \SpecCpWall\% of the energy of the wall pressure and \SpecUt\% of that of the first-row velocity.
The cutoff along the surface is therefore dictated by the near-wall flow; the outer flow alone would be represented
accurately by far fewer modes.

In the wall-normal direction a grid column is not periodic, so its spectrum is computed with the cosine transform, which
introduces no artificial discontinuity at the ends of the column. The energy beyond $k_\eta=11$ is only \EtaCp\% for the
pressure, \EtaUt\% for the velocity and about 1\% for the scaled velocity and the effective viscosity
[Fig.~\ref{fig:spectra}(b)]. This rapid decay is a consequence of the grid stretching. The boundary-layer profile varies
over five orders of magnitude in wall distance, but the geometric clustering of the rows makes its change between
consecutive rows nearly uniform, so that in index space the profile is a smooth function. \ours{} therefore retains
$-16\le k_\xi\le15$ and the smaller set $0\le k_\eta\le11$.

The cutoff restricts the information exchanged by the global path, not the spatial resolution of the output. A field truncated to
these modes could not represent the wall pressure better than about $\sqrt{0.055}\approx23\%$, yet the surface-pressure error
of \ours{} is \HoCp\% (Sec.~\ref{sec:res_held}). The local path, the pointwise nonlinearity, which turns smooth inputs into
sharper outputs because the product of two Fourier modes contains the sum of their wavenumbers, and the pointwise decoder are not
band-limited, and together they supply the finer features.

\subsubsection{Conditioning on the flow parameters by feature-wise linear modulation}\label{sec:film_method}

The Reynolds number and the angle of attack affect the whole flow at once. Besides entering as inputs at every node, they
therefore modulate every layer through feature-wise linear modulation (FiLM).\cite{perez2018film} In FiLM, an auxiliary
network computes from the conditioning variables a scale and a shift for each channel of the intermediate features of the
main network. Here a small network (2--32--512) maps $(Re,\alpha)$ to the channel-wise scale $\gamma_l$ and shift $\beta_l$
of Eq.~(\ref{eq:layer}). Its last layer is
initialized to zero, so that training starts from an unmodulated operator.

\subsubsection{Output encoding and wall factor}\label{sec:wall_method}

The decoder returns a scaled velocity $\hat{\mathbf u}=(\hat u,\hat v)$, $\log\nueff$, and the pressure in the form
$\tilde p=\operatorname{arcsinh}[(p-p_\infty)/\rho\Uinf^2]$, which is linear near the free-stream value and logarithmic at
the suction peak, where $C_p$ reaches $-11$; the logarithm of $\nueff$ plays the same role for a quantity that varies over three
orders of magnitude. The velocity needs more care, because of the way the wall shear is obtained from it.

\begin{figure*}[!tbp]
\centering
\includegraphics[width=\textwidth]{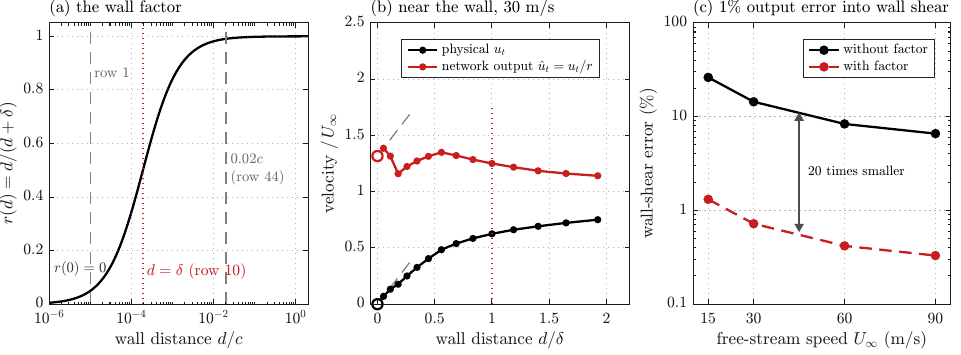}
\caption{\label{fig:wallfactor}The wall factor, on the reference solution of the held-out NACA~2415 at $8^\circ$, suction
side at $x/c=0.30$. (a) The factor of Eq.~(\ref{eq:wallfactor}): zero at the wall, linear below $\delta$, and 1 in the
outer flow; $\delta$ is the height of grid row 10. (b) Tangential velocity next to the wall at 30\,m/s, on linear axes
(dots: grid rows; dotted: $d=\delta$): the physical velocity, which is zero at the wall (no-slip), and the value the network
has to output, $\hat u_t=u_t/r$, which tends to the finite value $\tau_w\delta/\mu$, proportional to the wall shear (open
circles: values at the wall); grey dashed: linear growth with the slope $\tau_w/\mu$ of the reference wall shear.
(c) Error of the wall shear caused by an error of 1\% of the free-stream speed in the network output near the wall, at the
four database speeds (logarithmic axis; the error grows in proportion to the output error): without the factor the shear is
taken from the first-row velocity, with it from the output divided by $d_1+\delta$, Eq.~(\ref{eq:conditioning}).}
\end{figure*}

The wall shear is the slope of the tangential velocity at the wall, $\tau_w=\mu\,\partial u_t/\partial y_n$, and on the grid it
is taken from the first row, $\tau_w=\mu u_{t,1}/d_1$. This is exact only if the velocity still grows linearly with distance
at the first row, $u_t\approx(\tau_w/\mu)\,d$, and with $d_1=10^{-5}c$ it does in every case: in wall units the first row
lies at $d^+=0.6$--3.2 at the four speeds, inside the viscous sublayer,\cite{pope2000,schlichting2017} and the velocity there
sits on the line of slope $\tau_w/\mu$ [Fig.~\ref{fig:wallfactor}(b)]. The first-row velocity is, however, a small number:
\WfShareA, \WfShareB{} and \WfShareD\% of the velocity at $0.02c$ at 15, 30 and 90\,m/s. A network whose error is a modest
fraction of the velocity everywhere therefore makes a large relative error there, and the error passes undiminished into
$\tau_w$: an error of 1\% of the free-stream speed becomes an error of 14\% in the wall shear at 30\,m/s and of 26\% at
15\,m/s [solid line in Fig.~\ref{fig:wallfactor}(c)]. No-slip itself, moreover, would have to be learned from the few
sensors that happen to lie on the wall row.

We remove both difficulties by building the known behavior of the velocity at the wall into the output, following the
practice of enforcing boundary conditions in neural solvers by multiplying the network output with a function that vanishes
on the boundary.\cite{lagaris1998,sukumar2022} \ours{} multiplies its velocity by a \emph{wall factor},
\begin{equation}
\begin{gathered}
\mathbf u=r(d)\,\hat{\mathbf u},\qquad r(d)=\frac{d}{d+\delta},\\
\hat u_t=\frac{u_t}{r(d)}\approx\frac{\tau_w}{\mu}(d+\delta)\;\xrightarrow{\;d\to0\;}\;\frac{\tau_w\,\delta}{\mu},
\end{gathered}
\label{eq:wallfactor}
\end{equation}
where the second relation uses the linear growth of the velocity next to the wall. The factor is zero at the wall, so no-slip
holds exactly for any network output; it grows in proportion to the wall distance below $\delta$, as the velocity itself does;
and beyond $0.02c$ it differs from unity by less than 1\%, so the outer flow is unaffected [Fig.~\ref{fig:wallfactor}(a)].
The length $\delta=1.9\times10^{-4}c$, equal to the height of grid row 10, is constant: it does not change with the flow speed and assumes no shape of
the velocity profile, so the factor is a boundary constraint and not a wall function.

With the wall factor, the near-wall output of the network no longer has to vanish at the wall. It tends
instead to the finite value $\tau_w\delta/\mu$, which is proportional to the wall shear, while the physical velocity, $r\hat u_t$, remains
zero at the wall. Figure~\ref{fig:wallfactor}(b) shows both on the reference profile at 30\,m/s: the physical velocity
starts from zero at the wall and grows along the line of slope $\tau_w/\mu$, whereas the network output starts from the
finite value $\tau_w\delta/\mu$, here $1.3\,U_\infty$, and varies little over the first grid rows. From the first row to
$0.02c$ the physical velocity rises \WfRiseD- to \WfRiseA-fold at the four speeds, but the network output changes by at most a
factor of \WfScaledVar. The wall shear is then computed as $\mu\hat u_{t,1}/(d_1+\delta)$,
so an error $\epsilon$ in the network output changes the shear by $\mu\epsilon/(d_1+\delta)$ instead of $\mu\epsilon/d_1$,
\begin{equation}
\frac{\Delta\tau_w\big|_{\text{with }r}}{\Delta\tau_w\big|_{\text{without}}}=\frac{d_1}{d_1+\delta}\approx0.05,
\label{eq:conditioning}
\end{equation}
a reduction fixed by the grid alone and the same at every Reynolds number: the output error of 1\% of the free-stream speed, which
caused an error of 14\% in the wall shear at 30\,m/s, now causes an error of 0.7\% [dashed line in Fig.~\ref{fig:wallfactor}(c)]. The output is not smoother in index space, because
dividing by $r$ magnifies small irregularities of the reference in the first rows [Fig.~\ref{fig:spectra}(b)]; the benefit of
the factor lies in this reduced sensitivity of the wall shear, not in smoothness. The value of $\delta$ balances the two
requirements: it exceeds $d_1$ twentyfold, so the reduction is large, and it lies well inside the boundary layer, so the outer
flow is unaffected.

\subsubsection{Learning biases: sparse supervision and physics residuals}\label{sec:supervision}

A \emph{learning bias} imposes an assumption about the solution through the loss rather than through the structure of the
model: a penalty grows when the prediction violates the assumption, so that training favors solutions that satisfy it, but
nothing guarantees that the trained model does.\cite{karniadakis2021} Physics-informed neural networks are the best-known
example, with the residual of the governing equations added to the loss.\cite{raissi2019} A learning bias is softer than an
inductive one, and it has a property that matters here: it can act where there are no data, because a residual can be
evaluated at every node. Its weight must, however, be balanced against the data, and the data must themselves satisfy the
assumption, or the penalty will pull the model away from them.

The network predicts all 20\,480 nodes, but the data are known at only the 1024 sensors of each case. The loss therefore
compares prediction and reference only where the sensor mask $S$ of the case is one, and adds a residual of continuity on
the interior nodes $\Omega$,
\begin{equation}
\mathcal L=\frac{1}{4|S|}\sum_{i\in S}\sum_{k=1}^{4}\Bigl(\frac{\hat q_{k,i}-q_{k,i}}{\sigma_k}\Bigr)^{2}+\frac{\lambda}{|\Omega|}\sum_{j\in\Omega}r_{c,j}^{2},
\label{eq:loss}
\end{equation}
where $\hat q_k$ and $q_k$ are the predicted and reference raw outputs ($\hat u$, $\hat v$, $\tilde p$, $\log\nueff$),
$\sigma_k$ their standard deviation over the training sensors, and $\lambda=0.1$; the wall row is excluded for the
velocity, where $\hat{\mathbf u}$ is undetermined.

Mass is conserved at every node, observed or not, and checking this needs no data. Continuity is imposed through the relative
residual
\begin{equation}
r_c=\frac{\partial_xu+\partial_yv}{|\partial_xu|+|\partial_yv|+\kappa},
\label{eq:cont}
\end{equation}
evaluated with second-order differences and the chain rule of the grid. Velocity gradients in the boundary layer exceed those
of the outer flow by orders of magnitude, so an absolute residual would act almost entirely there; the relative form weights
the wall region and the far field alike, and $\kappa=0.05$ keeps it bounded in uniform flow. The interior nodes $\Omega$
exclude the wall row, the two outermost rows, and the few grid columns next to the sharp trailing edge (5\% of the
columns), where the mapping of the grid is singular. Such a penalty is beneficial only if the reference data
themselves nearly satisfy the constraint; for continuity this holds, the mean squared residual of the reference fields being
0.0037.

The momentum balance is the natural next constraint, and it is examined as an optional addition, normalized in the same way,
\begin{equation}
r_x=\frac{a_x+b_x-c_x}{|a_x|+|b_x|+|c_x|+\kappa},
\label{eq:mom}
\end{equation}
with the advection $a_x=(\mathbf u\cdot\nabla)u$, the pressure gradient $b_x=\partial_xp/\rho$ and the stress
$c_x=[\nabla\cdot\bm\tau]_x$, $\bm\tau=\nueff(\nabla\mathbf u+\nabla\mathbf u^{\mathsf T})$, and $r_y$ analogous. Unlike
continuity, the data do not satisfy this residual everywhere. Over all interior nodes the mean squared momentum residual of the
reference is 32 times its continuity residual, because the solver's finite-volume discretization and our finite differences on
the interpolated field disagree most where the stress is a second derivative taken across a few rows, that is, near the wall.
Only beyond $\dmin=10^{-3}c$ are the two residuals comparable, and the term is therefore admitted there alone: when it is used,
$w$ times the mean of $\tfrac12(r_x^2+r_y^2)$ over the nodes with $d\ge\dmin$ is added to Eq.~(\ref{eq:loss}).

\subsubsection{Derived wall quantities and training}\label{sec:training}

No surface quantity is regressed. All follow from the predicted field through the discrete counterparts of
Eq.~(\ref{eq:wallq}): $C_p$ on the wall row, $\tau_w=\mu u_{t,1}/d_1$ from the first row, $C_L$ by trapezoidal integration,
and $\xsep$ as the first sign change of the suction-side $\tau_w$ between $0.02c$ and $0.98c$. Deriving rather than
regressing ties these quantities to the field, so a model cannot report a plausible $C_f$ from an implausible near-wall
profile. \ours{} is trained with Adam\cite{kingma2015} for 6000 epochs (learning rate $10^{-3}$ with cosine decay to
$10^{-5}$, batches of four cases), in \TrainMin\,min on one graphics processing unit (GPU). The held-out score is recorded every 100 epochs, and the
state with the lowest score, at epoch 5600, is the one reported; every model and variant of this paper is selected by the
same rule. The score reaches a plateau after about 3000 epochs (Appendix~\ref{app:training}), so the choice of epoch changes
little, but the held-out conditions are thereby not fully independent of training; the new airfoils, evaluated
against OpenFOAM, are not used for any selection and therefore constitute the independent test set. The parameters
$\lambda$, $\kappa$ and $\delta$ were fixed a priori by the arguments above and were not tuned.

%=====================================================================
\subsection{Evaluation protocol}\label{sec:protocol}
%=====================================================================

A statement that one bias matters more than another is meaningful only if the operators compared received the same
information and the same optimization. Every model in this study therefore shares the training cases, the sensors, the
grid, the inputs, the raw outputs, the loss of Eq.~(\ref{eq:loss}), the optimizer, the 6000-epoch budget and the evaluation
code, and differs from \ours{} in one bias only (Table~\ref{tab:ingredients}). The modes, the width and the depth are
changed in a sweep of 1000 epochs each because of their cost. \ours{} and its two spectral variants are trained with two
seeds, which differ by \HoCpPm{} percentage points about their mean in held-out $C_p$ and \HoCfPm{} in $C_f$; differences
of that size are treated as ties.

Field errors are measured as relative $\ell_2$ errors weighted by the cell area $A_j$,
\begin{equation}
e(q)=\Bigl(\sum_jA_j(\hat q_j-q_j)^2\Bigr)^{1/2}\Big/\Bigl(\sum_jA_jq_j^2\Bigr)^{1/2},
\label{eq:l2}
\end{equation}
taken over the grid, over a band of wall distance, or over $d<0.02c$ for the near-wall velocity $u_\mathrm{w}$. A
field-averaged error is dominated by the large outer flow, so the results are also shown as \emph{error fields}, $|\hat q-q|$
at every node, which locate an error in the flow. At the wall, the surface-pressure error is the relative $\ell_2$ error of
$C_p$ over the wall, the lift error is $|\hat C_L-C_L|/\max(|C_L|,0.05)$, and the $C_f$ error of a case is the median
relative error of $\tau_w$ where the reference shear exceeds 5\% of its maximum.

The prediction of the separation point is quantified by the coefficient of determination between the predicted and the
reference $\xsep$ over a set of cases,
\begin{equation}
R^2_\mathrm{sep}=1-\frac{\sum_n\bigl(\hat x_{\mathrm{sep},n}-x_{\mathrm{sep},n}\bigr)^2}{\sum_n\bigl(x_{\mathrm{sep},n}-\overline{x}_\mathrm{sep}\bigr)^2},
\label{eq:r2sep}
\end{equation}
where a suction side that stays attached is assigned $\xsep=c$, separation at the trailing edge. A missed or a false
separation then enters the score as a location error rather than being discarded. $R^2_\mathrm{sep}=1$ is a perfect
prediction, $R^2_\mathrm{sep}=0$ is no better than predicting the mean, and negative values are worse than that. The score
is evaluated over all 96 database cases, of which 50 separate, because the eight held-out cases are too few to rank models,
and, on the new airfoils, over the \CondUnN{} solutions at unseen conditions, where shape, angle and speed are all new.

A model that observes 5\% of each field can also reproduce what it observes without learning what lies between, as a
regression can pass through its points without generalizing. The training fields contain a test of their own, the 95\% of
their nodes that the loss never sees, and we measure it with the memorization ratio
\begin{equation}
M=\frac{e_{\bar S}(u)}{e_S(u)},
\label{eq:memo}
\end{equation}
where $e_S$ and $e_{\bar S}$ are the relative root-mean-square errors of $u$ over the sensor nodes and over all other nodes
of the same training case, averaged over the 88 cases. A model that has learned the field has $M$ near 1, whereas one that
reproduces its observations far better than the field between them has $M\gg1$. Every model trained to fit its sensors has
$M>1$, \ours{} included ($M=\HoMemo$), and a model that fits its sensors poorly can have a small $M$ without learning more of
the field, so $M$ is used to compare versions of one model and is always reported together with the absolute error at the
unobserved nodes.

On the new airfoils, errors are measured against OpenFOAM. In incompressible flow the pressure is determined only up to an
additive constant, which each solver fixes through its far-field boundary treatment. Pressure errors against OpenFOAM are
therefore evaluated \emph{gauge-free}: the prediction is first shifted by the area-weighted mean difference,
\begin{equation}
\hat p\to\hat p-\frac{\sum_jA_j(\hat p_j-p_j)}{\sum_jA_j}.
\label{eq:gauge}
\end{equation}

\subsection{Comparison architectures}\label{sec:baselines}

To assess whether the Fourier backbone itself matters, it is replaced by seven other architectures that retain the grid,
the inputs, the outputs, the wall factor, the sensors, the loss and the training of \ours{}, so that the comparison is fair
in the sense of Refs.~\onlinecite{lu2022fair,mcgreivy2024}. The architectures were selected to span the principal ways in
which neural surrogates exchange information between nodes, and they include the networks most widely used for airfoil
flows. Ordered from global to pointwise exchange, they are:
(1) Transolver,\cite{wu2024transolver} a transformer that applies attention among learned slices of the domain;
(2) the deep operator network (DeepONet),\cite{lu2021deeponet} which expands each output field in learned basis functions;
(3) the U-Net,\cite{ronneberger2015} the standard convolutional encoder--decoder for airfoil RANS
fields,\cite{thuerey2020} which combines local operations at several resolutions;
(4) a residual convolutional network (ResNet),\cite{he2016} with local convolutions only;
(5) MeshGraphNet,\cite{pfaff2021} a graph neural network that passes messages between neighboring nodes;
(6) PointNet,\cite{qi2017} a point-cloud network that adds one global feature to pointwise layers; and
(7) a multilayer perceptron (MLP), which acts pointwise and exchanges no information between nodes.
Their sizes and training times are given in Appendix~\ref{app:models}; each is trained with one seed.

Two non-learned references complete the comparison. The first is $k$-nearest-neighbor regression on all training sensors.
The second is sensor interpolation, which triangulates the 1024 sensors of the test case itself. Sensor interpolation thus
uses observations of the test case that no other model receives, and it indicates how much of the field the sparse
observations of one case recover without any learning.

%=====================================================================
\section{Results and discussion}\label{sec:results}\label{sec:complete}
%=====================================================================

This section first evaluates the complete operator, at held-out conditions of the training airfoils (Sec. III A) and on unseen airfoils against an independent solver (Sec. III B). The contribution of each bias is then isolated by removing or altering it, first for the inductive biases (Sec. III C) and then for the learning biases (Sec. III D). Section III E ranks these effects, and Sec. III F compares GP-FNO with seven other architectures in accuracy and computational cost. The results show that the wall factor and the continuity residual have the largest effects, on the wall quantities and on the fields between the sensors, respectively.

\subsection{Prediction at held-out flow conditions}\label{sec:res_held}

\begin{figure*}[!tbp]
\centering
\includegraphics[width=\textwidth]{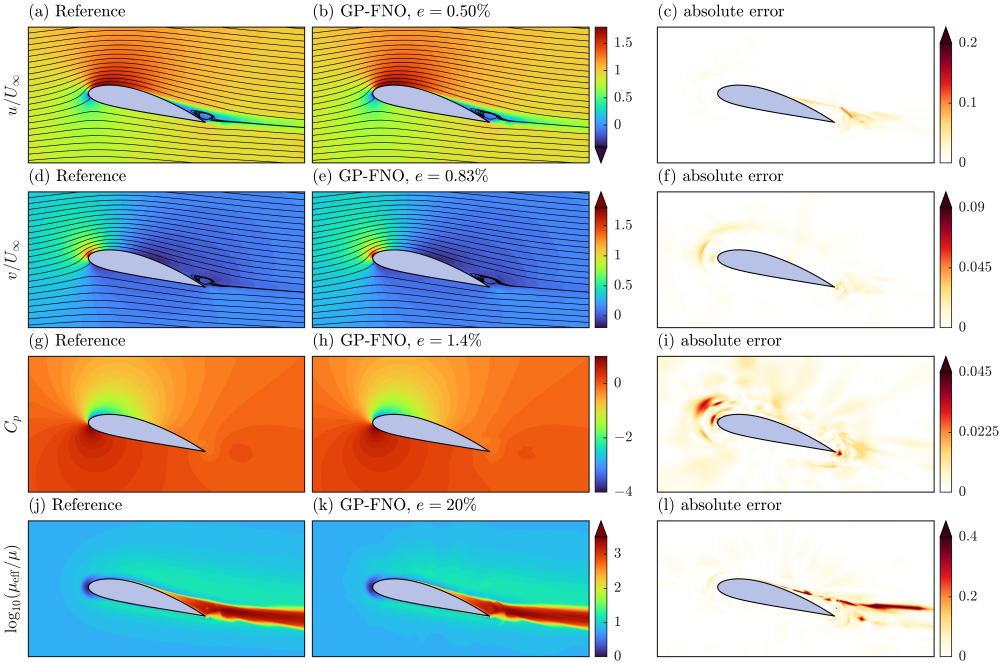}
\caption{\label{fig:heldfields}The complete operator on the separated held-out NACA~4418 at $\alpha=14^\circ$, 30\,m/s:
reference field, prediction of \ours{} and absolute error, for the two velocity components, the pressure coefficient and
the effective viscosity. The percentage in each prediction panel is the area-weighted relative error, Eq.~(\ref{eq:l2}).}
\end{figure*}

Figure~\ref{fig:heldfields} shows the most demanding held-out condition, NACA~4418 at $14^\circ$ and 30\,m/s, whose
boundary layer separates at $x/c=0.70$. Neither the condition nor any field near it was seen in training, and the
neighboring conditions were observed at only 5\% of their nodes. The prediction reproduces the acceleration over the
leading edge, the stagnation region below it, the low-velocity region over the rear of the suction side, and the wake.
The error fields localize the discrepancies. In velocity [Fig.~\ref{fig:heldfields}(c, f)] the error is
confined to the shear layer that leaves the separation point and to the near wake, where a small displacement of a thin
layer appears as a large local error. In pressure [Fig.~\ref{fig:heldfields}(i)] it forms faint rings around the leading
edge and the trailing edge, the signature of a slightly misplaced suction peak. The effective viscosity is the least
accurate field [Fig.~\ref{fig:heldfields}(l)], with its error in the near wake, where $\mu_\mathrm{eff}/\mu$ reaches
$10^3$. Averaged
over the eight held-out conditions, the field errors are \HoU\% in $u$, \HoP\% in $p$ and \HoNu\% in $\nueff$.

\begin{figure*}[!tbp]
\centering
\includegraphics[width=\textwidth]{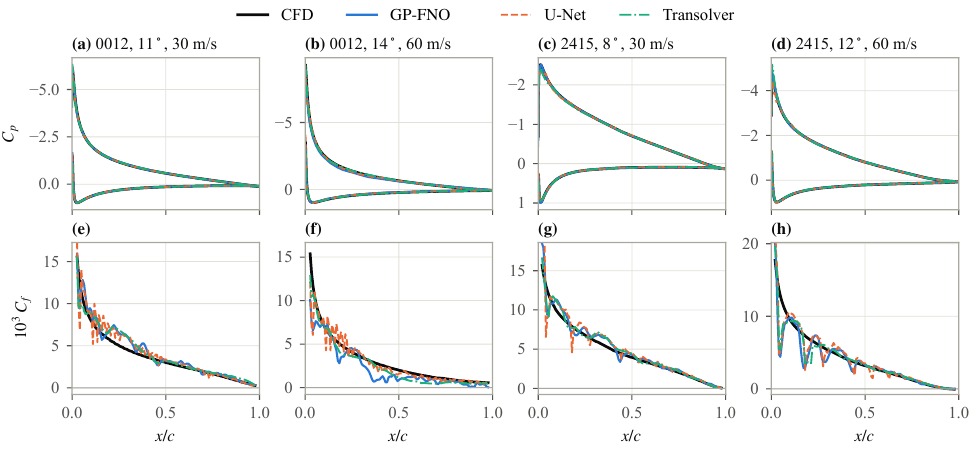}
\caption{\label{fig:heldsurf}Surface pressure (upper row) and suction-side skin friction (lower row) on four held-out
conditions, derived from the predicted fields of \ours{}, the U-Net and Transolver; CFD: the reference field. Panels
(b, f) show the largest error of \ours{} in the held-out set.}
\end{figure*}

The wall quantities are a stricter test, because they are never supervised (Fig.~\ref{fig:heldsurf}; every held-out case
is listed in Table~\ref{tab:heldout} of Appendix~\ref{app:cases}). The surface pressure is indistinguishable from the
reference on the scale of the figure except near the steepest suction peaks, and the lift is within 0.7\% on seven of the
eight conditions. The mean errors are \HoCp\% in $C_p$ and \HoCl\% in $C_L$, and the predicted separation points of the
held-out cases lie within $0.01c$ of the reference. The derived skin friction follows the reference with a median error of
\HoCf\%, although no wall shear was ever supervised. The largest error occurs for NACA~0012 at $14^\circ$ and 60\,m/s, which has
the steepest suction peak of the held-out set ($C_p\approx-9$), at \HoClMax\% in lift [Fig.~\ref{fig:heldsurf}(b, f)]: the predicted peak
is lower than the reference, and the case lies two degrees below the angle at which the lift of this airfoil collapses at
the same speed in the training data. The U-Net also has its largest surface-pressure error on this case (5.3\%), whereas
Transolver, which attends over the whole domain, resolves the peak better (2.7\%). On the other seven conditions the surface-pressure error
of \ours{} stays between 1.0 and 3.0\%, that of the U-Net between 1.0 and 4.1\% and that of Transolver between 1.9 and
6.5\%.

\subsection{Generalization to unseen airfoils and an independent solver}\label{sec:res_new}

\begin{table*}[!t]
\caption{\label{tab:newair}Generalization to new airfoils, measured against OpenFOAM. Medians over the cases of each row;
pressure errors are gauge-free, Eq.~(\ref{eq:gauge}). The first row involves no model: it is the difference between the
two RANS solvers on the held-out conditions, the code gap.}
\begin{ruledtabular}
\begin{tabular}{lrrrrrr}
 & Cases & $u$ (\%) & $p$ (\%) & $C_p$ (\%) & $C_L$ (\%) & $C_f$ (\%) \\
\hline
Code gap: Fluent against OpenFOAM, held out & 8 & 0.50 & 5.3 & 3.7 & 2.8 & 15.1 \\
GP-FNO, held out & 8 & 0.65 & 5.2 & 3.9 & 2.0 & 17.4 \\
\hline
GP-FNO, new airfoils, trained conditions & 64 & 0.54 & 6.3 & 5.7 & 3.7 & 13.0 \\
GP-FNO, new airfoils, unseen conditions & 72 & 0.56 & 5.2 & 5.1 & 2.9 & 11.5 \\
GP-FNO, all new airfoils & 136 & 0.56 & 5.9 & 5.3 & 3.2 & 12.0 \\
\end{tabular}
\end{ruledtabular}
\end{table*}

The operator was trained on Fluent solutions only, and on the new airfoils it is evaluated against OpenFOAM, the
independent solver of Sec.~\ref{sec:reference}. Table~\ref{tab:newair} summarizes the \Nnew{} solutions. Its first row gives
the code gap, the median difference between the two solvers at the eight held-out conditions, which involves no learned
model: \GapU\% in $u$, \GapP\% in $p$, \GapCp\% in $C_p$, \GapCl\% in $C_L$ and \GapCf\% in $C_f$. The corresponding median
errors of \ours{} on the unseen airfoils are \NewU\%, \NewP\%, \NewCp\%, \NewCl\% and \NewCf\%.

For the velocity, the pressure field and the skin friction, the errors of \ours{} are therefore comparable to or smaller
than the code gap. For the surface pressure and the lift they are larger, and the excess quantifies the additional error
caused by the unseen shape. A substantial part of the measured error thus originates from the difference between the
solvers rather than from the operator. The variation of the code gap between the training airfoils supports this
interpretation: its value in $C_p$ is \GapCpA\% for NACA~0012, \GapCpB\% for NACA~2415 and \GapCpC\% for NACA~4418, and the
$C_p$ error of \ours{} on each new airfoil follows the code gap of its nearest training airfoils, with a correlation
coefficient of \GapCorr{} across the eight shapes.

\begin{figure*}[!tbp]
\centering
\includegraphics[width=\textwidth]{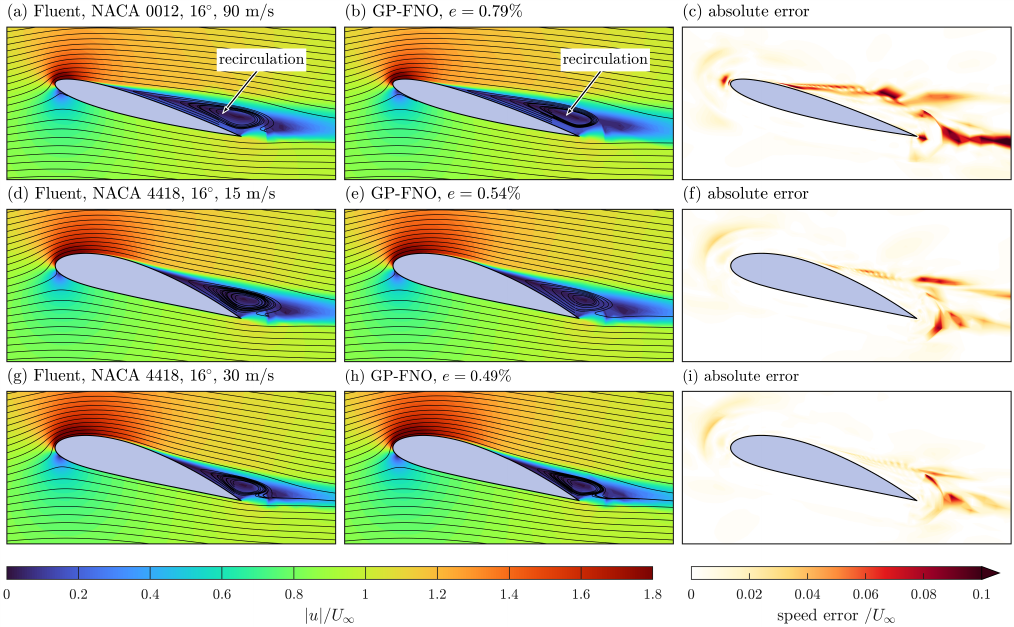}
\caption{\label{fig:recirc}Recirculation on three training conditions at $\alpha=16^\circ$, drawn with the free stream
horizontal: Fluent solution, prediction of \ours{} with streamlines, and absolute error of the speed; the arrows mark the
closed eddy. The percentage in each prediction panel is the area-weighted relative error of $u$, Eq.~(\ref{eq:l2}).
NACA~0012 at 90\,m/s and NACA~4418 at 15\,m/s are past maximum lift.}
\end{figure*}

\begin{figure*}[!tbp]
\centering
\includegraphics[width=\textwidth]{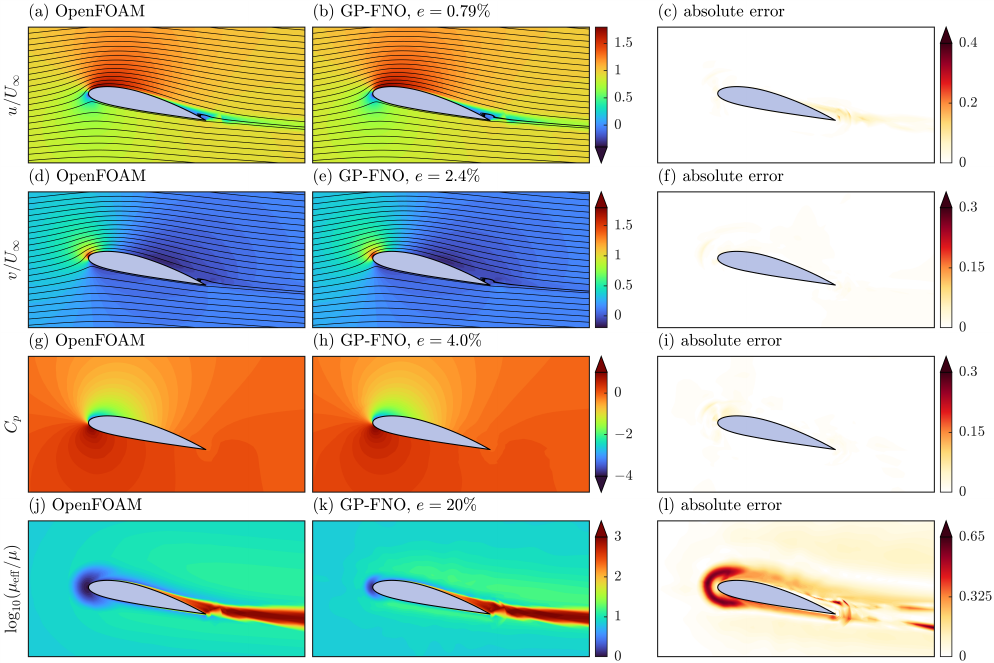}
\caption{\label{fig:newstall}A new airfoil at an unseen condition: NACA~3416, $\alpha=13^\circ$, 20\,m/s, near stall.
OpenFOAM solution, prediction of \ours{} and absolute error; $C_p$ of the prediction is on the OpenFOAM gauge,
Eq.~(\ref{eq:gauge}). The percentage in each prediction panel is the area-weighted relative error, Eq.~(\ref{eq:l2}). Each
error field has the same color scale in Figs.~\ref{fig:newstall}, \ref{fig:newhighspeed} and \ref{fig:newlargest}.}
\end{figure*}

Figure~\ref{fig:newstall} shows a new airfoil at an unseen condition, where shape, angle and speed are all new at once:
NACA~3416 at $13^\circ$ and 20\,m/s, close to stall with reversed flow near the trailing edge. Its error panels share one
color scale with the two further cases of Appendix~\ref{app:extra}. The velocity and pressure fields are reproduced with
errors close to the code gap, and the error concentrates at the leading edge and in the near wake, as it did on the
held-out case of Fig.~\ref{fig:heldfields}. The largest error is again in the effective viscosity, whose logarithm
$\log_{10}(\mu_\mathrm{eff}/\mu)$ is in error by up to about 0.5 in a band around the airfoil and in the wake; this is also the
field in which the two codes differ most (code gap
\GapNu\%), so that panel is not a measure of the model alone.

\begin{figure*}[!tbp]
\centering
\includegraphics[width=\textwidth]{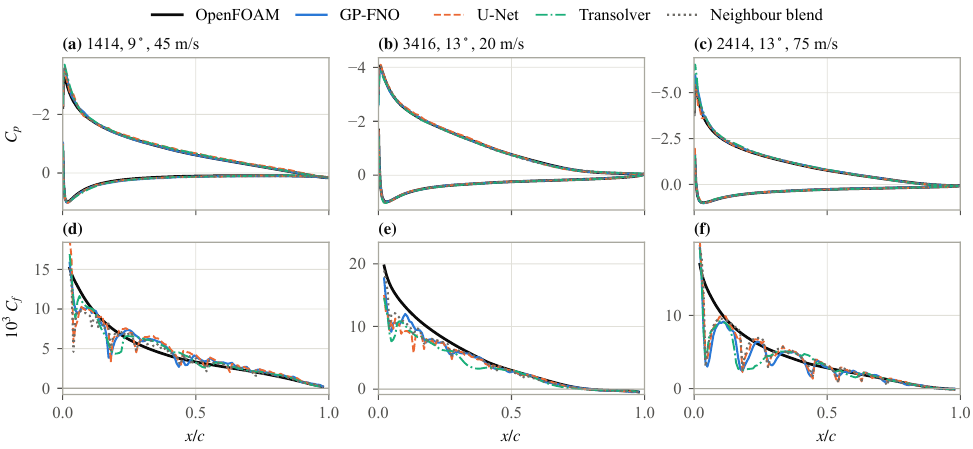}
\caption{\label{fig:newsurf}Surface pressure and suction-side skin friction derived from the predicted fields on three new
airfoils at unseen conditions, against OpenFOAM's wall values. Predicted $C_p$ is on the OpenFOAM gauge. Neighbour blend: a
non-learned baseline that interpolates the reference solutions of the bracketing trained conditions (linearly in $\alpha$
and in $\log\Uinf$) on the two training airfoils nearest in shape, weights them by camber and maps the result to the new
surface. Panels (c, f) show the case of median $C_p$ error over all new-airfoil solutions.}
\end{figure*}

Figure~\ref{fig:newsurf} gives the wall quantities of this case, of NACA~1414 at $9^\circ$ and 45\,m/s, attached on the
suction side, and of NACA~2414 at $13^\circ$ and 75\,m/s, the case
of median surface-pressure error over all new-airfoil solutions, with the U-Net, Transolver and a non-learned blend of the
neighboring training solutions for comparison. The surface pressure is reproduced closely on all three, with $C_p$ errors
of 6.6, 3.4 and 5.4\%. Near the leading edge the skin friction of every learned model lies below OpenFOAM's, most clearly at
75\,m/s [Fig.~\ref{fig:newsurf}(f)], and farther downstream the curves oscillate about it. In this region the models agree
more closely with one another than with OpenFOAM, so the underprediction is attributable to the training target and to the
code gap rather than to a particular architecture. Two further cases, at a high speed and at the edge of the angle range, are shown with their wall
quantities in Appendix~\ref{app:extra}.

\begin{figure*}[!tbp]
\centering
\includegraphics[width=\textwidth]{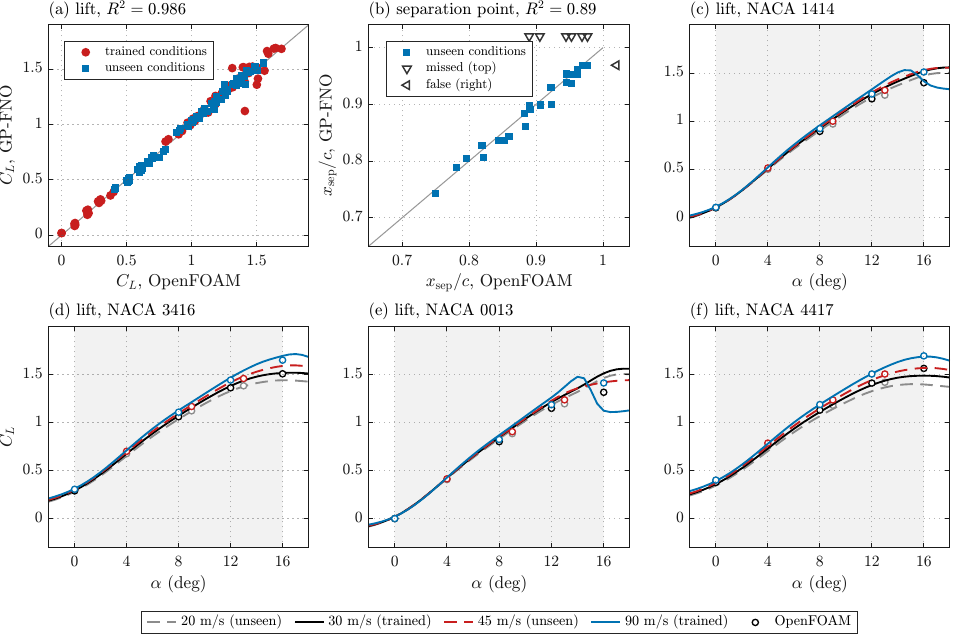}
\caption{\label{fig:newsummary}\ours{} on the OpenFOAM solutions of the eight new airfoils. (a) Lift against OpenFOAM on all
\Nnew{} solutions, at trained and unseen conditions, with the coefficient of determination. (b) Separation point on the
\CondUnN{} solutions at unseen conditions, where both codes find one, with $R^2_\mathrm{sep}$ of Eq.~(\ref{eq:r2sep});
triangles: missed separation (top edge) or spurious separation (right edge). (c--f) Lift curves of four new airfoils, \ours{} swept
continuously in angle (lines; dashed at the unseen speeds) and OpenFOAM (circles); shaded: training range of angle.}
\end{figure*}

\begin{table}[!htbp]
\caption{\label{tab:newangle}\ours{} on the \Nnew{} new-airfoil solutions, by angle of attack: medians and maxima of
the gauge-free surface-pressure error and of the absolute lift error, the median skin-friction error, and the
coefficient of determination of the separation point, Eq.~(\ref{eq:r2sep}), given at the unseen conditions only.
Trained angles are flown at 30 and 90\,m/s, unseen angles at 20, 45 and 75\,m/s.}
\begin{ruledtabular}
\begin{tabular}{rlrrrrrrr}
 & & & \multicolumn{2}{c}{$C_p$ (\%)} & \multicolumn{2}{c}{$|\Delta C_L|$} & $C_f$ (\%) & \\
\cline{4-5}\cline{6-7}
$\alpha$ ($^\circ$) & Set & Cases & med. & max. & med. & max. & med. & $R^2_\mathrm{sep}$ \\
\hline
0 & trained & 16 & 5.8 & 9.3 & 0.016 & 0.030 & 13.3 &  \\
4 & unseen & 24 & 5.4 & 8.5 & 0.018 & 0.052 & 10.0 & 0.71 \\
8 & trained & 16 & 4.3 & 9.9 & 0.025 & 0.047 & 11.3 &  \\
9 & unseen & 24 & 3.9 & 10.0 & 0.021 & 0.047 & 11.3 & 0.92 \\
12 & trained & 16 & 6.4 & 15.0 & 0.036 & 0.075 & 12.4 &  \\
13 & unseen & 24 & 5.2 & 15.1 & 0.042 & 0.101 & 12.8 & 0.76 \\
16 & trained & 16 & 8.6 & 22.9 & 0.076 & 0.288 & 15.9 &  \\
\hline
all & & 136 & 5.3 & 22.9 & 0.026 & 0.288 & 12.0 & 0.89 \\
\end{tabular}
\end{ruledtabular}
\end{table}

Figure~\ref{fig:newsummary} and Table~\ref{tab:newangle} show the new-airfoil cases at once. The lift has a median absolute
error of \LiftMedAbs{} and a coefficient of determination of \NewClR{} [Fig.~\ref{fig:newsummary}(a)]. The errors depend on
the angle more than on whether the condition was trained (Table~\ref{tab:newangle}): from $0^\circ$ to $13^\circ$ the median
surface-pressure error stays between 3.9 and 6.4\% and the median lift error below 0.05, and both are largest at
$16^\circ$, where the flow approaches stall and the reference itself is least certain. The separation point is evaluated at the
unseen conditions, where shape, angle and speed are all new: there $R^2_\mathrm{sep}=\HoNewSepR$, between \AngSepRLo{} and
\AngSepRHi{} at the three unseen angles, and where both solvers predict separation the median absolute error of the separation location is
\SepLocUn$c$ [Fig.~\ref{fig:newsummary}(b)]. The predicted lift curves agree with the OpenFOAM values at trained and unseen
speeds alike [Fig.~\ref{fig:newsummary}(c--f)]. The exception is NACA~0013 at 90\,m/s, whose predicted lift collapses above $14^\circ$
while OpenFOAM keeps lifting [Fig.~\ref{fig:newsummary}(e)]: the model inherits the early stall of its symmetric training
neighbor, NACA~0012, whose Fluent solution at 90\,m/s stalls between $14^\circ$ and $16^\circ$, three degrees before the
experiment [Fig.~\ref{fig:validation}(f)]. Figure~\ref{fig:recirc} shows this stall and two separated NACA~4418 cases at
$16^\circ$, training conditions observed at only 5\% of their nodes: \ours{} reproduces the closed eddy over the rear of
the suction side, with reversed-flow areas of 0.040, 0.035 and 0.024$c^2$ against 0.044, 0.037 and 0.024$c^2$ in Fluent,
and errors of 0.5--0.8\% in $u$.

%=====================================================================
\subsection{Role of the inductive biases}\label{sec:inductive}\label{sec:unpack}

This subsection and the next isolate the effect of each bias by removing or altering it while every other element of the
protocol is held fixed. For each bias we examine the change in the predicted flow and in the wall quantities, the
mechanism inside the trained network, and whether the effect persists on the unseen airfoils; Table~\ref{tab:variants}
(Sec.~\ref{sec:synthesis}) collects the numbers. The inductive biases of Secs.~\ref{sec:geometry}--\ref{sec:wall_method}
are examined first, in the order in which information passes through the operator. The body-fitted grid is shared by every
model and is not removed; the other biases are altered or removed, and the operator is also applied to grids on which it was
not trained.

%---------------------------------------------------------------------
\subsubsection{The global spectral path}\label{sec:spectral}
%---------------------------------------------------------------------

The argument of Sec.~\ref{sec:paths} suggests that the global path carries the pressure coupling of the flow. We tested this
with two variants. In the first, the Fourier transform is applied along $\xi$ only, so that every grid row is convolved along
the surface with the same learned kernel and no global coupling acts in the wall-normal direction. In the second, the global
path is removed altogether, which leaves four local layers with FiLM and the wall factor.

\begin{figure*}[!tbp]
\centering
\includegraphics[width=\textwidth]{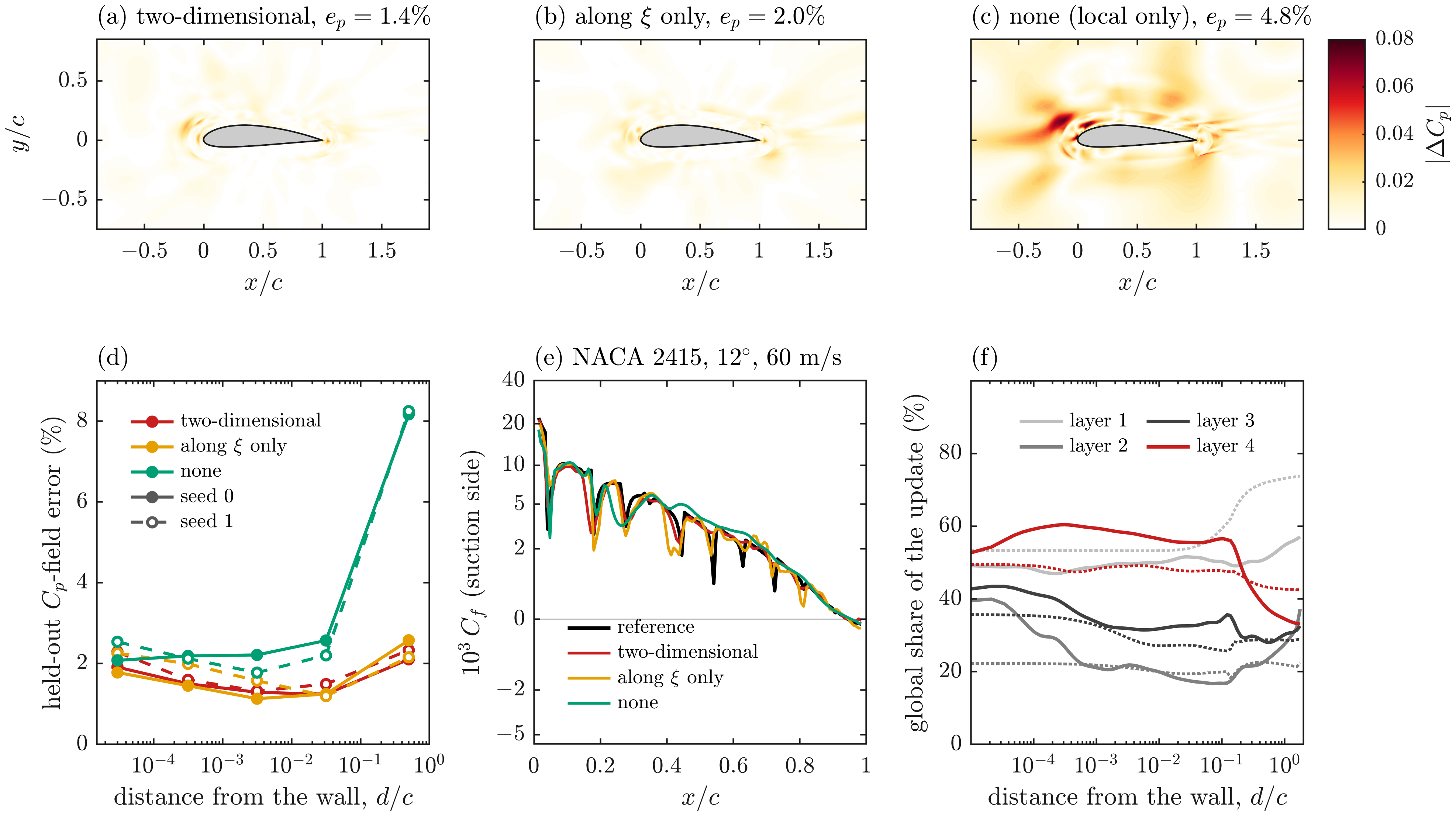}
\caption{\label{fig:spectral}The global spectral path. (a--c) Absolute error of the pressure coefficient in the field of the
held-out NACA~4418 at $14^\circ$, 30\,m/s, for two-dimensional Fourier layers (\ours{}), layers along $\xi$ only and no
global path, with the relative pressure error of the case. (d) Held-out pressure error by band of wall distance, both
seeds. (e) Suction-side skin friction of the held-out NACA~2415 at $12^\circ$, 60\,m/s. (f) Share of the global path in
the update of each layer of \ours{} (solid) and of the variant along $\xi$ (dotted), against wall distance, averaged over
the held-out conditions.}
\end{figure*}

The pressure error fields of Fig.~\ref{fig:spectral}(a--c) illustrate the difference. With two-dimensional layers the
error is confined to a faint ring around the leading edge, and the variant along $\xi$ shows a similar pattern. Without a global path,
however, it spreads over the whole domain, with lobes of error upstream of the leading edge, above and below the airfoil and
far into the outer flow. Resolved by wall distance [Fig.~\ref{fig:spectral}(d)], the pressure error of the local network
rises from \HoBandPE\% to \LocBandPE\% beyond $0.1c$, for both seeds, while within $0.02c$ of the wall it stays within a
percentage point of \ours{}. This is the region in which, according to Eq.~(\ref{eq:poisson}), the pressure is determined by
the flow elsewhere in the domain, and a stack of four local layers, whose receptive field spans only seven nodes, cannot
transmit that information.

At the wall the result is less intuitive. The surface pressure does not require the global path: the local network
reproduces the held-out $C_p$ as well as \ours{} ($\LocCpMean\pm\LocCpPm$\% against $\HoCpMean\pm\HoCpPm$\%), because every
node receives $(Re,\alpha)$ and its own geometry directly, and on these airfoils the wall pressure is, to a good
approximation, a function of these. With only three training airfoils, whose camber and thickness vary together, the
surface pressure can largely be interpolated from these local inputs, and a broader variation of shape may be required
before the global path improves the surface pressure. The global path does, however, affect the wall shear and the
separation point. The local network reaches $\LocCfMean\pm\LocCfPm$\% in $C_f$ and $R^2_\mathrm{sep}=\,$\LocSepRMean~$\pm$~\LocSepRPm, against
$\HoCfMean\pm\HoCfPm$\% and \HoSepRMean~$\pm$~\HoSepRPm{} for \ours{}. The variant along $\xi$ has the worst skin friction of
any variant that keeps the wall factor, $\XiCfMean\pm\XiCfPm$\%, with $R^2_\mathrm{sep}=\,$\XiSepRMean~$\pm$~\XiSepRPm, and
its predicted $C_f$ exhibits spurious oscillations [Fig.~\ref{fig:spectral}(e)]. A plausible explanation is that the single
kernel along $\xi$ is shared by all grid rows. It is therefore adapted mainly to the outer rows, whose energy lies at low
$k_\xi$, and is ill suited to the wall row, whose spectrum extends to higher wavenumbers (Fig.~\ref{fig:spectra}).

The contribution of the global path in the trained operator is measured by its share in the update of each layer,
$\|G_l\|/(\|G_l\|+\|L_l\|)$ [Fig.~\ref{fig:spectral}(f)]. Neither path dominates. The global path supplies about half of the
update in the first layer and a quarter to a third in the next two. In the last layer, which forms the output, it supplies
\GlobLDWall\% of the update within the boundary layer and \GlobLDOut\% far from it, so the global coupling is strongest where
the near-wall output is formed. the global path determines the pressure field and makes the wall quantities consistent with it. Both require coupling in the wall-normal direction
as well as along the surface. The same pattern is observed on the new airfoils (Table~\ref{tab:variants}).

%---------------------------------------------------------------------
\subsubsection{Retained modes, width and depth}\label{sec:capacity}
%---------------------------------------------------------------------

\begin{figure*}[!tbp]
\centering
\includegraphics[width=\textwidth]{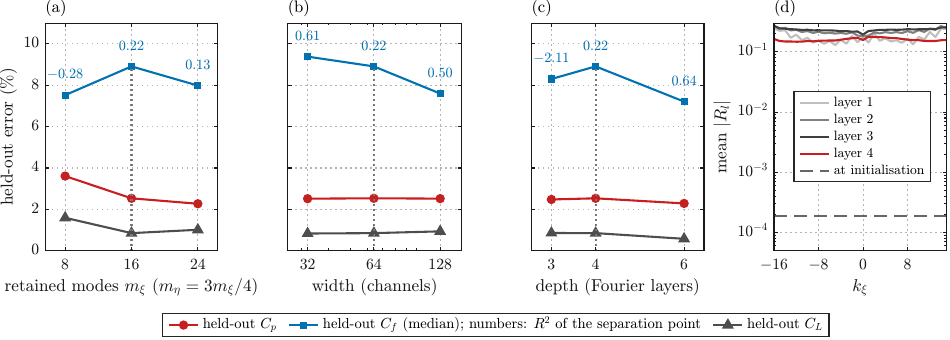}
\caption{\label{fig:capacity}Spectral resolution and capacity. (a--c) Held-out errors after 1000 epochs of training against
the retained modes, the width and the depth; dotted line: adopted configuration; numbers above the $C_f$ curve:
$R^2_\mathrm{sep}$ of the separation point over the 96 database cases, Eq.~(\ref{eq:r2sep}). (d) Mean magnitude of the
learned spectral weights $R_l$ of \ours{} at each wavenumber along the surface, averaged over $k_\eta$; dashed: mean
magnitude at initialization.}
\end{figure*}

The spectra of Fig.~\ref{fig:spectra} set the retained modes in advance, and the sweep of Fig.~\ref{fig:capacity}(a--c)
checks that choice and the capacity around it (numbers in Table~\ref{tab:sweep}, Appendix~\ref{app:extra}). Halving the
retained modes in each direction, a cutoff that discards 16\% of the wall-pressure energy, raises the held-out $C_p$ error
from \SwBaseCp\% to \SwMACp\%, nearly doubles the lift error and lowers $R^2_\mathrm{sep}$ from \SwBaseSepR{} to \SwMASepR.
Raising the modes by half, on the other hand, improves $C_p$ only to \SwMCCp\%, at \SwMCPar{} million parameters against
\SwBasePar{} million. The spectra anticipate both outcomes, since the halved cutoff falls where the wall spectrum still
carries energy and the raised one where little is left. The trained weights agree [Fig.~\ref{fig:capacity}(d)]: at every
retained wavenumber and in every layer the mean magnitude of $R_l$ lies \WRatioMin--\WRatioMax{} times above its initial
value and shows no decay toward the cutoff. The trained operator thus uses every retained mode, and the information carried
by the global path is limited by the cutoff rather than by the learned weights.

Capacity matters less. For widths from 32 to 128 channels the surface-pressure error remains at \SwWACp\%, while $C_f$ improves
from \SwWACf\% to \SwWCCf\% at sixteen times the parameters. Depth acts mainly on separation, with $R^2_\mathrm{sep}=\,$\SwDASepR{}
for three layers against \SwBaseSepR{} and \SwDCSepR{} for four and six. After a sixth of the training the separation point
is not yet converged, so these values serve to compare configurations rather than to rate them; for the adopted
configuration, $R^2_\mathrm{sep}$ is \SwBaseSepR{} after 1000 epochs and \HoSepR{} after the full 6000. This configuration is
the smallest on the plateau of surface-pressure error, and the halved modes and the shallower network are the only changes
that markedly degrade the prediction of separation.

%---------------------------------------------------------------------
\subsubsection{Evaluation on refined grids}\label{sec:resolution}
%---------------------------------------------------------------------

\begin{figure*}[!tbp]
\centering
\includegraphics[width=\textwidth]{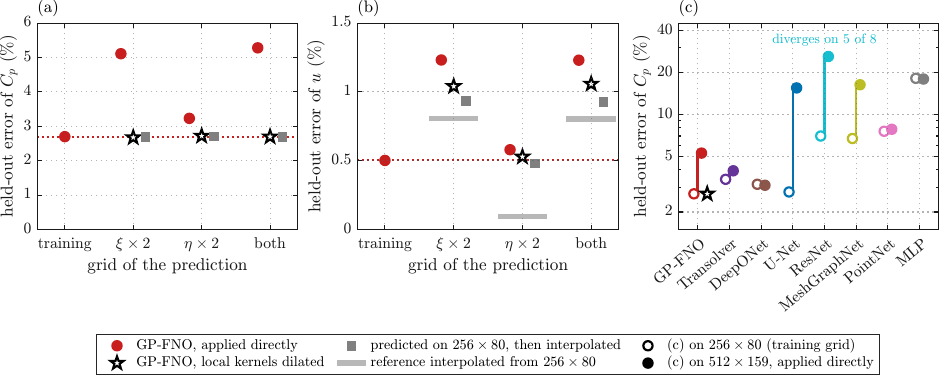}
\caption{\label{fig:refinement}The trained operators on refined grids, without retraining (held-out means). (a) Surface-pressure
and (b) velocity error of \ours{} on grids refined twice along $\xi$, twice along $\eta$, and both, applied directly, with
its $3\times3$ local kernels dilated by the refinement factor, and predicted on the training grid and interpolated; dotted:
error on the training grid; gray bars in (b): the reference itself interpolated from the training grid. (c) Surface-pressure
error of the eight architectures on the training grid and on the $512\times159$ grid, applied directly; star: \ours{} with
dilated kernels.}
\end{figure*}

A Fourier layer is defined by its weights on wavenumbers, not on grid points, so an FNO can in principle be evaluated on a
finer grid than the one it was trained on, and such ``zero-shot super-resolution'' is often counted among its
advantages.\cite{li2021fno,kovachki2023} We examined this property for \ours{} by applying the trained operator, without
retraining, to nested grids built by the same meshing rule with twice the columns ($512\times80$), twice the rows
($256\times159$), or both, with the zero padding along $\eta$ scaled accordingly.

Applied directly, \ours{} became less accurate on the finer grids [Fig.~\ref{fig:refinement}(a)]: its surface-pressure
error rose from \RfTrainCp\% to \RfBothDirCp\% on $512\times159$, and most of the rise came with the refinement along the
surface (\RfXiDirCp\%, against \RfEtaDirCp\% along $\eta$). The observations point to the local path rather than to the
spectral path. The spectral weights act on wavenumbers and transfer with the grid, whereas the $3\times3$ stencils act on
neighboring nodes, which lie closer on a finer grid, so that a stencil that learned a difference over $0.008c$ along the
surface is applied over $0.004c$. Consistent with this, dilating each local kernel by the refinement factor, so that it again
reaches the neighbors seen in training, returned the surface-pressure error to \RfBothDilCp\% on every refined grid without
retraining, and the velocity behaved alike [Fig.~\ref{fig:refinement}(b)].

The finer grid did not add accuracy in these tests. With dilated kernels \ours{} was as accurate as when it predicted on the
training grid and its prediction was interpolated (\RfBothIntCp\% in $C_p$), and slightly less accurate in velocity
(\RfBothDilU\% against \RfBothIntU\%), both close to the error of interpolating the reference itself (\RfFloorU\%). Among the
other architectures [Fig.~\ref{fig:refinement}(c)], those built from local stencils lost more, the U-Net reaching
\RfBothUnCp\% and MeshGraphNet \RfBothMgCp\% in $C_p$ and the residual network (ResNet) diverging on five of the eight
cases, whereas Transolver, the deep operator network (DeepONet), PointNet and the multilayer perceptron (MLP), which act on coordinates rather than on neighbors, kept their training-grid
accuracy. In this test, then, the loss of accuracy of \ours{} on refined grids came from its local path, dilating the local
kernels recovered it, and the gain we observed is consistency across grids rather than super-resolution.

%---------------------------------------------------------------------
\subsubsection{Conditioning by feature-wise linear modulation (FiLM)}\label{sec:film}
%---------------------------------------------------------------------

\begin{figure*}[!tbp]
\centering
\includegraphics[width=\textwidth]{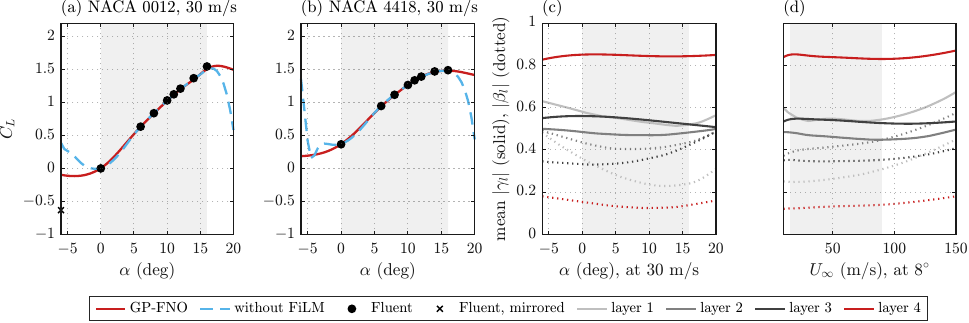}
\caption{\label{fig:film}Conditioning on the flow parameters. (a, b) Lift at 30\,m/s beyond the training angles (shaded:
training range) for \ours{} and the variant without FiLM, with the Fluent solutions and, for the symmetric NACA~0012, the
Fluent solutions mirrored to negative angles. (c, d) Mean magnitude of the learned scale $\gamma_l$ (solid) and shift
$\beta_l$ (dotted) of each layer: (c) against angle at 30\,m/s and (d) against speed at $8^\circ$; shaded: training ranges.}
\end{figure*}

Within the training ranges, FiLM does not improve the predictions. Without it the held-out errors are equal or lower in every
quantity except the skin friction ($C_p$ \NfCp\% against \HoCp\%, $C_f$ \NfCf\% against \HoCf\%), and on the new airfoils
nothing changes. The trained modulation explains why [Fig.~\ref{fig:film}(c, d)]: the channels are rescaled by up to 85\%,
yet over the training angles the scale varies by at most \GamVarA{} and over the training speeds by at most \GamVarU. FiLM
has learned a nearly constant per-channel gain, which a network without it can absorb into its weights, and the dependence on
$(Re,\alpha)$ enters mainly through the inputs.

Its effect appears only in extrapolation [Fig.~\ref{fig:film}(a, b)]. Without FiLM the lift of NACA~0012 turns upward again
below $0^\circ$ and collapses beyond $16^\circ$ ($C_L=\NfClTwenty$ at $20^\circ$), whereas with FiLM it stays monotonic and
bounded, because the modulation continues its trends smoothly. Both are wrong below $0^\circ$, where the symmetric airfoil
must give $C_L=-0.63$ at $-6^\circ$, since nothing in the inputs or the losses tells the operator that the airfoil is
symmetric. Parameter conditioning therefore keeps extrapolation bounded, but it does not improve accuracy inside the data.

%---------------------------------------------------------------------
\subsubsection{The wall factor}\label{sec:wall}
%---------------------------------------------------------------------

\begin{figure*}[!tbp]
\centering
\includegraphics[width=\textwidth]{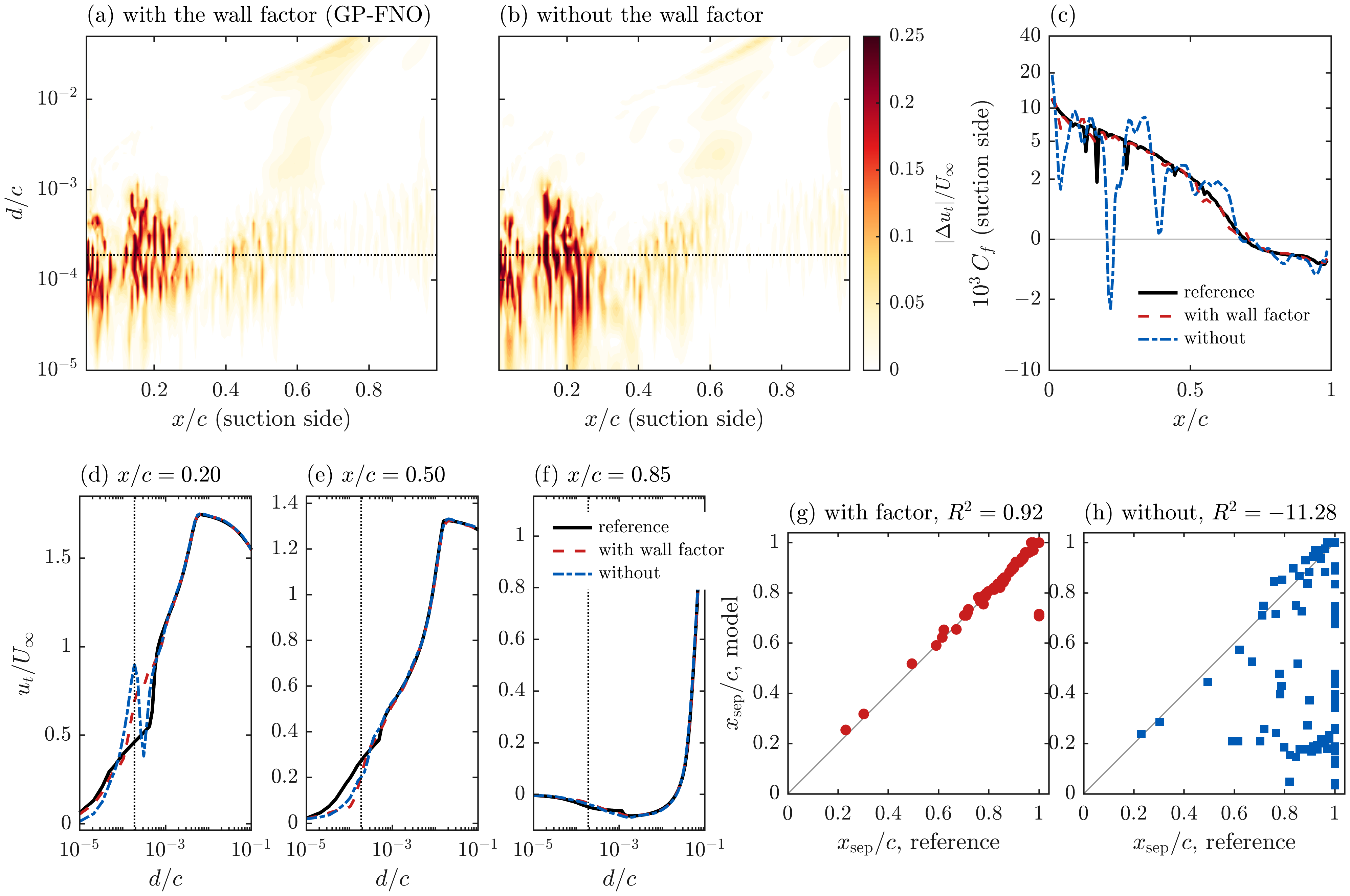}
\caption{\label{fig:wall}The wall factor, on the separated held-out NACA~4418 at $14^\circ$, 30\,m/s. (a, b) Absolute error
of the tangential velocity on the suction side in boundary-layer coordinates, streamwise position against wall distance
(logarithmic), with and without the wall factor; dotted line: $d=\delta$. (c) Suction-side skin friction and (d--f)
tangential-velocity profiles at three stations, dotted line $d=\delta$: reference (solid), with (dashed) and without
(dash-dotted) the wall factor; the flow separates at $x/c=0.70$. (g, h) Predicted against reference separation point over
the 96 database cases, with and without the wall factor, and $R^2_\mathrm{sep}$ of Eq.~(\ref{eq:r2sep}); attached flow is
placed at $\xsep/c=1$, so a missed separation lies on the top edge and a false one on the right edge.}
\end{figure*}

When the wall factor is removed, the held-out $C_p$ and $C_L$ remain as good as with it (\NwCp\% and \NwCl\%), but the
skin-friction prediction deteriorates sharply: its error rises from \HoCf\% to \NwCf\%, fivefold, and the coefficient of determination of the
separation point over the 96 database cases falls from \HoSepR{} to \NwSepR.

The error fields of Fig.~\ref{fig:wall}(a, b) show where this happens. They are drawn with the wall distance on a logarithmic
axis, so that both the viscous sublayer and the outer part of the boundary layer are resolved. Both operators make their largest errors in the front third of the suction side, below $d=10^{-3}c$, where the
boundary layer is thinnest, but without the factor these errors are stronger and reach down to the wall row, where the factor
holds the velocity of \ours{} at zero. At $x/c=0.2$ the operator without the factor overshoots the reference profile near
$d=\delta$ and then dips below it [Fig.~\ref{fig:wall}(d)], while farther aft both operators follow the profile closely,
including the reversed flow at $x/c=0.85$.

This near-wall error degrades the predicted wall quantities. Without the factor, the first-row velocity, from which $\tau_w$ is
computed, carries a ripple along the surface, so the predicted $C_f$ oscillates about the reference and, at $x/c\approx0.2$,
changes sign [Fig.~\ref{fig:wall}(c)]. Such a spurious sign change is read as separation. Over the 96 database cases the
operator without the factor therefore reports separation at $\xsep\approx0.15$--$0.3c$ for many cases that separate near the
trailing edge, and falsely for many attached ones [Fig.~\ref{fig:wall}(h)], whereas \ours{} lies on the diagonal
[Fig.~\ref{fig:wall}(g)]. The energy of the first-row velocity error beyond the retained modes is \RippleHigh{} times that of
\ours{}: the error without the factor is a short-wavelength ripple, which is precisely the scale that produces sign changes
of the wall shear.

The factor acts only where it should. It lowers the held-out velocity error below $d=10^{-4}c$ from \NwBandUA\% to
\HoBandUA\%, leaves it unchanged between $10^{-4}c$ and $10^{-3}c$, and costs some accuracy farther out (\HoBandUC\% against
\NwBandUC\% between $10^{-3}c$ and $10^{-2}c$). The large effect on $C_f$ thus comes from the first few rows alone, as
Eq.~(\ref{eq:conditioning}) anticipates, and it carries over to the new airfoils, where the $C_f$ error without the factor is
\NwNewCf\% rather than \NewCf\% and $R^2_\mathrm{sep}$ at the unseen conditions falls from \HoNewSepR{} to \NwNewSepR. Building no-slip into the output
therefore decides whether skin friction and separation can be learned from sparse data at all.

\subsection{Role of the learning biases}\label{sec:learning}

The learning biases of Sec.~\ref{sec:supervision} act through the loss. Unlike the inductive biases they leave the
structure of the operator unchanged, and their effect is read in what the trained operator predicts at the nodes that
carry no data.

%---------------------------------------------------------------------
\subsubsection{The continuity residual}\label{sec:continuity}
%---------------------------------------------------------------------

\begin{figure*}[!tbp]
\centering
\includegraphics[width=\textwidth]{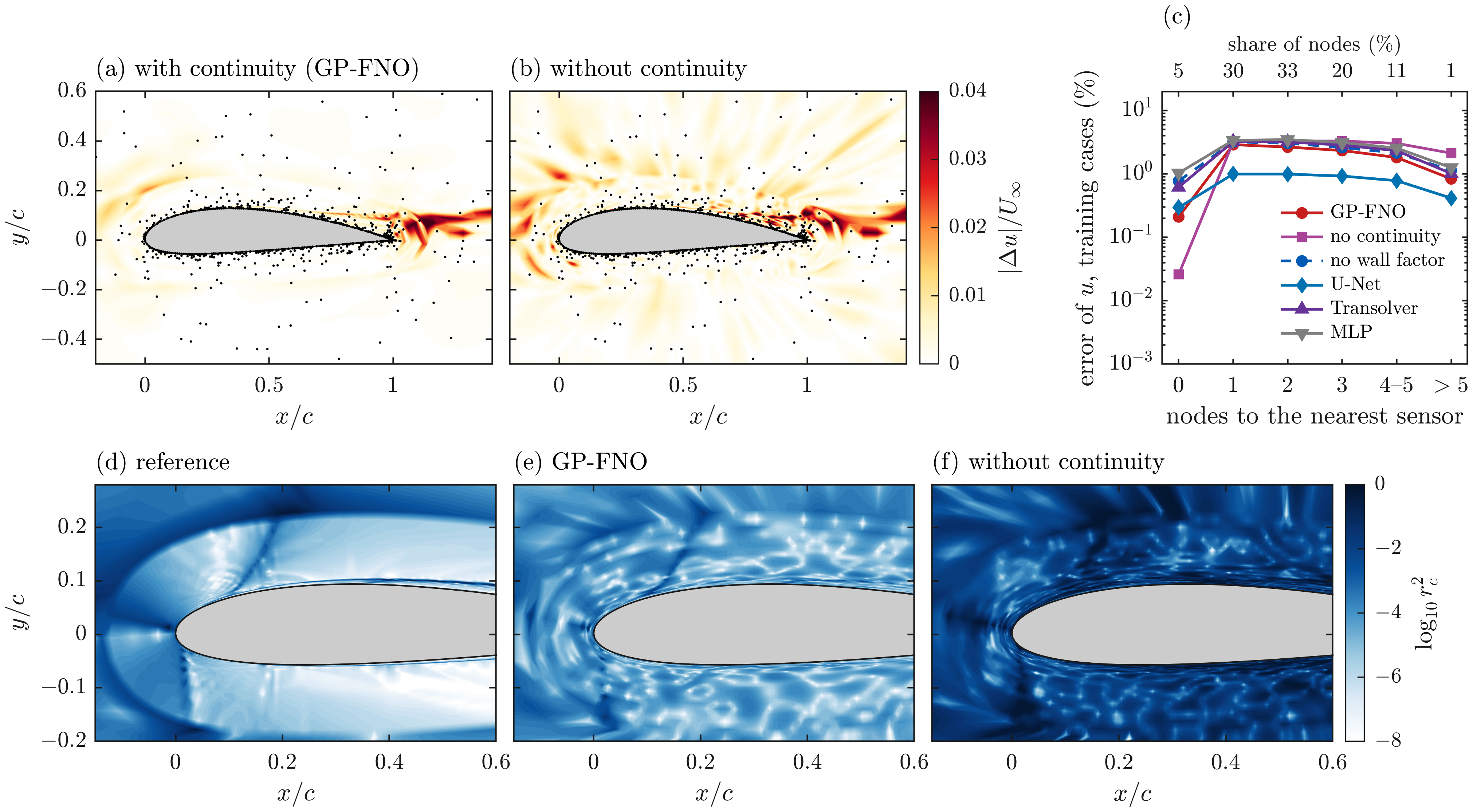}
\caption{\label{fig:continuity}The continuity residual. (a, b) Absolute error of $u$ on a training case, NACA~4418 at
$12^\circ$, 30\,m/s, with its 1024 sensors (dots), with and without the residual. (c) Error of $u$ over the 88 training
cases, grouped by the distance of each node from the nearest sensor of its case (grid nodes; distance 0: the sensors
themselves); top: share of the nodes in each group. (d--f) Squared continuity residual, Eq.~(\ref{eq:cont}), around the
front half of the held-out NACA~2415 at $12^\circ$, 60\,m/s: reference field, \ours{}, and \ours{} trained without the
residual.}
\end{figure*}

The continuity residual is the only part of the loss that sees the 95\% of each training field that carries no data. Its
effect is clearest on a training case, where the answer between the sensors is known but was never shown to the network
[Fig.~\ref{fig:continuity}(a, b)]. With the residual, the error is a smooth field concentrated in the wake and the shear
layer. Without it, the error forms a web of streaks between the sensors, while at the sensors themselves it nearly vanishes.

The node-wise errors quantify this behavior. Without the residual, \ours{} reproduces its sensors to \NcSens\% in $u$, while its error
at the unobserved nodes of the same fields is \NcGap\%, a memorization ratio of $M=\NcMemo$: the network has learned the
observed values rather than the flow between them. With the residual the sensor error rises to \HoSens\%, the unobserved
error falls to \HoGap\%, by 22\%, and $M=\HoMemo$; the residual thus reduces the memorization by an order of magnitude
without removing it. These node-wise errors are relative root-mean-square errors over nodes and are not comparable with the
area-weighted field errors of Sec.~\ref{sec:complete}. Figure~\ref{fig:continuity}(c) resolves this by the distance of each node from the
nearest sensor. Without continuity the error is \NcDistZero\% at the sensors and \NcDistOne--\NcDistThree\% one to three nodes
away, and it hardly falls with distance, whereas with continuity it is \HoDistZero\% at the sensors and falls to \HoDistFour\%
four to five nodes away. These distances are short: \ShareThree\% of all nodes lie within three nodes of a
sensor. The continuity-residual fields show the same contrast in terms of mass conservation around the leading edge, along the boundary
layer of the front half of the airfoil and in the flow outside it [Fig.~\ref{fig:continuity}(d--f)]. \ours{} keeps the
residual small everywhere in this region, with a median of $4.6\times10^{-5}$ against $4.7\times10^{-6}$ for the reference;
without the term the median is $3.2\times10^{-2}$, about 700 times that of \ours{}, and the imbalance is largest in the
boundary layer and around the nose. Over the whole field the median residual without the term is \ContRatio{} times that
of \ours{}.

The benefit reaches the held-out conditions and the new airfoils. Without the residual the held-out errors in $u$, $p$ and
$C_p$ rise to \NcU\%, \NcP\% and \NcCp\% (from \HoU\%, \HoP\% and \HoCp\%), and on the new airfoils the field error in $u$
rises from \NewU\% to \NcNewU\%. The wall shear and separation are hardly affected, because they are set by the first rows
and the wall factor. A conservation law that the data satisfy is therefore an effective constraint on the unobserved nodes,
provided it is imposed in a relative form that weights the wall region and the outer flow alike.

%---------------------------------------------------------------------
\subsubsection{The momentum residual}\label{sec:momentum}
%---------------------------------------------------------------------

\begin{figure*}[!tbp]
\centering
\includegraphics[width=\textwidth]{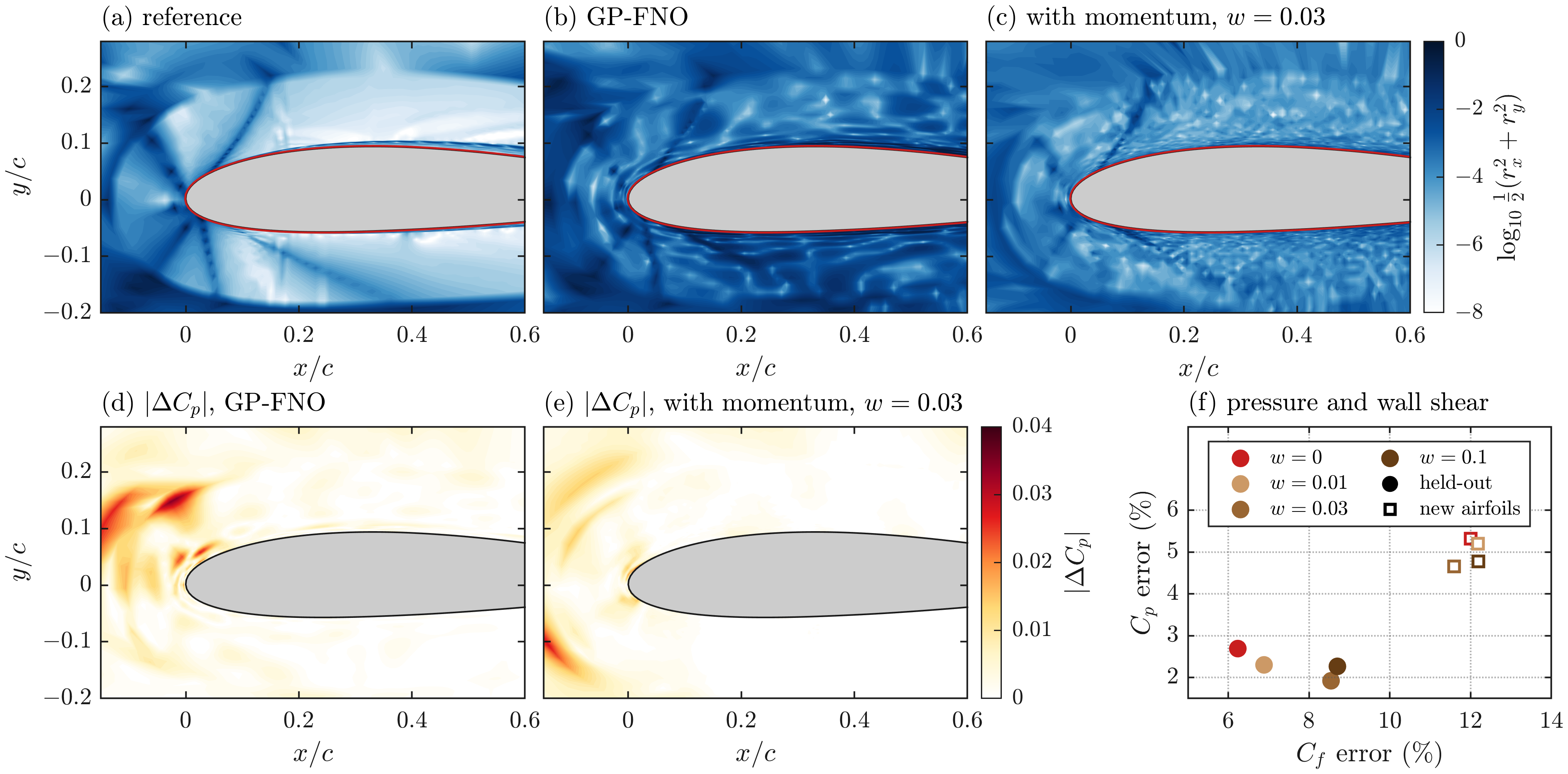}
\caption{\label{fig:momentum}The momentum residual around the front half of the held-out NACA~2415 at $12^\circ$,
60\,m/s. (a--c) Squared momentum residual $\tfrac12(r_x^2+r_y^2)$ of Eq.~(\ref{eq:mom}): reference field, \ours{}, and
\ours{} trained with the momentum term at $w=0.03$; red line: $d=\dmin=10^{-3}c$, inside which the term is not applied.
(d, e) Absolute error of $C_p$ in the field, without and with the term. (f) Surface-pressure error against skin-friction
error at the three weights, on the held-out conditions (mean $C_p$, median $C_f$) and on the new airfoils (medians).}
\end{figure*}

The momentum balance couples velocity and pressure, and its residual is therefore a natural candidate for a further
constraint. As shown in Sec.~\ref{sec:supervision}, the data satisfy it only beyond $\dmin$, so the term is applied there alone, at three
weights, and all three are reported, because selecting one by the held-out results would amount to selection on the test set.

The residual maps around the front half of the airfoil show what the term does to the flow
[Fig.~\ref{fig:momentum}(a--c)]. Beyond $\dmin$, in the region shown, \ours{} trained without the term has a median momentum
residual of $7.9\times10^{-3}$, nearly a thousand times the $9.4\times10^{-6}$ of the reference, in streaks over the
boundary layer and the outer flow; with $w=0.03$ it falls twentyfold, to $3.9\times10^{-4}$. Inside $\dmin$, where the term
is not applied, all three fields keep a large residual, the thin dark layer along the wall.

The effect on the errors is selective. The momentum term reduces the pressure error: at all three weights the held-out $C_p$ error
falls, from \HoCp\% to \MaCp\%, \MbCp\% and \McCp\%. Around the front half of the airfoil it removes the lobes of pressure
error above the leading edge, though a weaker one appears upstream of the stagnation point [Fig.~\ref{fig:momentum}(d, e)],
and the mean $|\Delta C_p|$ over the region shown falls by 29\%, from 0.0032 to 0.0023. The error falls inside $\dmin$ as
well, from 0.0046 to 0.0035, although the term acts only beyond it, because pressure is a global field and correcting it away from the wall corrects it next to the wall too. The velocity in
the first rows, which sets the wall shear, is unchanged or slightly worse (Table~\ref{tab:variants}), and the $C_f$ error
rises to \MaCf--\McCf\% [Fig.~\ref{fig:momentum}(f)]. On the
new airfoils the benefit persists and the cost disappears, with $C_p$ falling from \NewCp\% to \MbNewCp--\MaNewCp\% and $C_f$
staying at \MbNewCf--\MaNewCf\%. A physics residual should therefore be checked against the data before it is added: the
momentum balance, admissible only away from the wall, improves the pressure throughout the field, but not the wall shear,
which is determined by the first grid rows, where the term is not applied.

%=====================================================================
\subsection{Synthesis: ranking of the biases}\label{sec:synthesis}
%=====================================================================

\begin{table*}[!tp]
\caption{\label{tab:variants}Effect of every change to the operator. Held-out columns: means over the eight held-out
conditions against the reference field (median for $C_f$), in percent; $u_\mathrm{w}$ is the error of $u$ over
$d<0.02c$; $R^2_\mathrm{sep}$ is the coefficient of determination of the separation point over all 96 database cases,
Eq.~(\ref{eq:r2sep}); $M$ is the memorization ratio, Eq.~(\ref{eq:memo}). New-airfoil columns: medians over the \Nnew{}
OpenFOAM solutions, gauge-free, in percent, and $R^2_\mathrm{sep}$ over the \CondUnN{} solutions at unseen conditions.}
\setlength{\tabcolsep}{3.6pt}
\begin{ruledtabular}
\begin{tabular}{llrrrrrrrrrrrr}
 & & \multicolumn{8}{c}{Held-out conditions} & \multicolumn{4}{c}{New airfoils} \\
\cline{3-10}\cline{11-14}
Operator & & $u$ & $p$ & $u_\mathrm{w}$ & $C_p$ & $C_L$ & $C_f$ & $R^2_\mathrm{sep}$ & $M$ & $C_p$ & $C_L$ & $C_f$ & $R^2_\mathrm{sep}$ \\
\hline
GP-FNO & seed 0 & 0.50 & 1.9 & 5.6 & 2.7 & 1.1 & 6.2 & 0.92 & 12.8 & 5.3 & 3.2 & 12.0 & 0.89 \\
 & seed 1 & 0.51 & 2.2 & 6.1 & 3.0 & 1.5 & 6.0 & 0.80 & 12.8 & 5.0 & 3.1 & 12.0 & 0.93 \\
\hline
Global path along $\xi$ & seed 0 & 0.37 & 2.3 & 5.8 & 1.9 & 0.9 & 16.9 & $-$1.61 & 2.4 & 5.8 & 3.2 & 19.7 & $-$8.50 \\
 & seed 1 & 0.36 & 2.0 & 5.6 & 2.3 & 1.3 & 19.4 & $-$0.88 & 2.7 & 6.2 & 4.0 & 21.0 & $-$7.30 \\
No global path & seed 0 & 0.56 & 7.1 & 4.9 & 2.1 & 0.7 & 7.6 & $-$2.73 & 1.2 & 4.9 & 2.0 & 14.4 & $-$17.5 \\
 & seed 1 & 0.63 & 7.1 & 4.7 & 2.7 & 0.6 & 8.5 & $-$4.02 & 1.2 & 5.7 & 2.9 & 13.7 & $-$26.9 \\
\hline
No FiLM &  & 0.38 & 1.6 & 4.9 & 2.5 & 0.7 & 7.7 & 0.92 & 12.2 & 5.4 & 3.1 & 11.9 & 0.94 \\
No wall factor &  & 0.44 & 2.2 & 5.8 & 2.2 & 0.7 & 30.9 & $-$11.3 & 3.7 & 4.7 & 3.5 & 29.2 & $-$49.6 \\
No continuity &  & 0.77 & 3.8 & 5.6 & 3.9 & 1.1 & 6.8 & 0.75 & 149.3 & 6.2 & 3.1 & 11.9 & 0.87 \\
\hline
Momentum & $w=0.01$ & 0.54 & 1.8 & 5.9 & 2.3 & 1.1 & 6.9 & 0.80 & 11.7 & 5.2 & 3.3 & 12.2 & 0.93 \\
 & $w=0.03$ & 0.46 & 1.6 & 5.5 & 1.9 & 0.8 & 8.5 & 1.00 & 10.4 & 4.7 & 3.2 & 11.6 & 0.92 \\
 & $w=0.1$ & 0.49 & 2.2 & 5.9 & 2.3 & 1.1 & 8.7 & 0.35 & 8.6 & 4.8 & 3.0 & 12.2 & 0.90 \\
\end{tabular}
\end{ruledtabular}
\end{table*}

\begin{figure*}[!tbp]
\centering
\includegraphics[width=\textwidth]{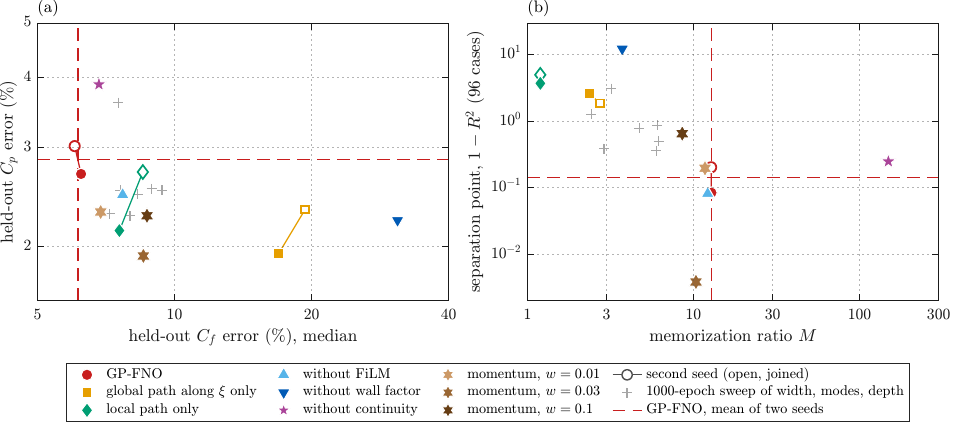}
\caption{\label{fig:tradeoff}Every change to the operator on one map. (a) Held-out surface-pressure error against
skin-friction error. (b) Unexplained share of the separation point over the 96 database cases, $1-R^2_\mathrm{sep}$ of
Eq.~(\ref{eq:r2sep}), against the memorization ratio $M$. One symbol per change; open symbols joined by a thin line: second
seed; crosses: the 1000-epoch sweep of width, modes and depth; dashed lines: \ours{}, mean of two seeds, which divide better
from worse in each quantity.}
\end{figure*}

\begin{figure*}[!tbp]
\centering
\includegraphics[width=\textwidth]{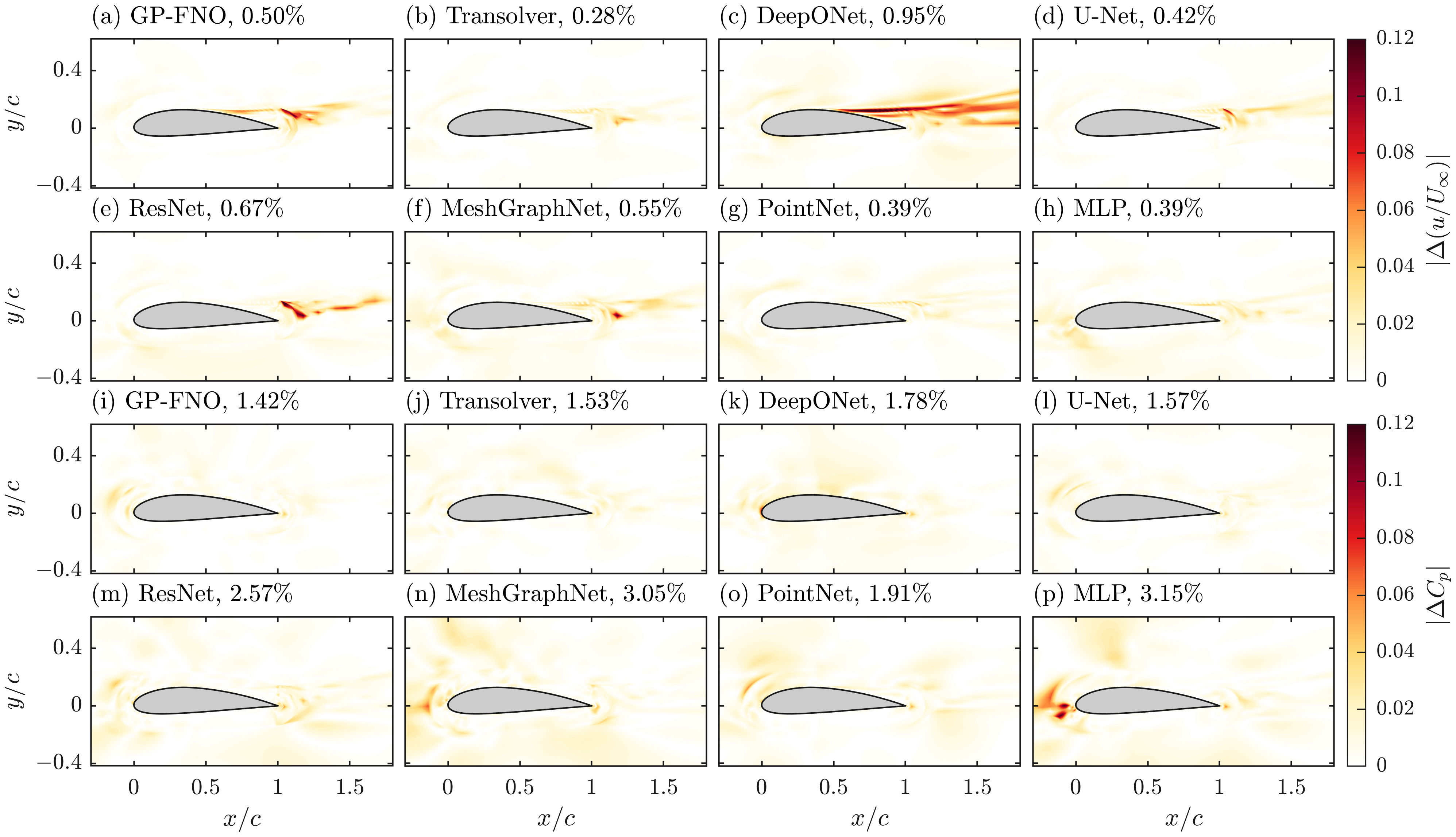}
\caption{\label{fig:bbfields}Error fields of the eight architectures on the separated held-out NACA~4418 at $14^\circ$,
30\,m/s, on one color scale: (a--h) streamwise velocity and (i--p) pressure coefficient, with the area-weighted relative
error of each field.}
\end{figure*}

Table~\ref{tab:variants} collects the effect of every change, and Fig.~\ref{fig:tradeoff} places the changed operators on
two maps. On the first, surface pressure against wall shear [Fig.~\ref{fig:tradeoff}(a)], removing the wall factor moves the
operator far to the right, a fivefold rise in skin-friction error with no loss in pressure, and restricting the global path
to $\xi$ moves it right by a factor of three; every other change moves it by less than a factor of two, mostly along the
pressure axis. On the second, separation against memorization [Fig.~\ref{fig:tradeoff}(b)], the removal of the continuity
residual stands alone at the far right, while the removal of the wall factor and the two spectral variants stand at the
top, where the separation point is predicted worse than by its mean ($R^2_\mathrm{sep}<0$). The crosses of the
width--modes--depth sweep fill the region between, which shows how much of the scatter is ordinary variation in capacity and
training.

Ranked by the magnitude of the change they cause, the biases fall into three groups. The first group determines what can
be learned at all. The wall factor determines whether the wall shear and the separation point can be learned. The
continuity residual determines whether the field between the sensors is learned.

The second group determines whether the predicted fields are mutually consistent. The two-dimensional global path makes the
pressure field consistent with the wall quantities. The retained modes, chosen from the wall spectrum, are all used
by the trained operator. In the refinement test, the local path accounted for the loss of accuracy on finer grids, and
dilating its kernels recovered it.

The third group only refines the result. The momentum residual improves the pressure, FiLM matters only in extrapolation,
and width and depth change little within the range of the training data.

Each effect appears on the new airfoils as on the held-out conditions (Table~\ref{tab:variants}), so the ranking is not
specific to the conditions on which it was measured. The roadmap of Table~\ref{tab:roadmap} (Sec.~\ref{sec:conclusions})
condenses these findings bias by bias.

%=====================================================================
\subsection{Comparison with other architectures}\label{sec:compare}
%=====================================================================

This subsection compares \ours{} with the seven architectures and the two non-learned references of
Sec.~\ref{sec:baselines}, all trained and evaluated under the same protocol.

\subsubsection{Predicted fields and wall quantities}\label{sec:compare_pred}

\begin{table*}[!tp]
\caption{\label{tab:backbones}GP-FNO and seven other architectures that share the grid, inputs, outputs, wall factor, sensors, loss and
training of GP-FNO, and two non-learned references. Columns as in Table~\ref{tab:variants}. Sensor interpolation
triangulates the sensors of the test case itself, information that no other model receives.}
\setlength{\tabcolsep}{3.4pt}
\begin{ruledtabular}
\begin{tabular}{llrrrrrrrrrrrr}
 & & \multicolumn{8}{c}{Held-out conditions} & \multicolumn{4}{c}{New airfoils} \\
\cline{3-10}\cline{11-14}
Model & Exchange between nodes & $u$ & $p$ & $u_\mathrm{w}$ & $C_p$ & $C_L$ & $C_f$ & $R^2_\mathrm{sep}$ & $M$ & $C_p$ & $C_L$ & $C_f$ & $R^2_\mathrm{sep}$ \\
\hline
GP-FNO & global, spectral & 0.50 & 1.9 & 5.6 & 2.7 & 1.1 & 6.2 & 0.92 & 12.8 & 5.3 & 3.2 & 12.0 & 0.89 \\
Transolver & global, attention & 0.24 & 1.6 & 5.5 & 3.4 & 0.9 & 6.7 & 1.00 & 5.0 & 6.3 & 3.6 & 14.1 & 0.92 \\
DeepONet & global, basis expansion & 0.84 & 2.4 & 5.9 & 3.2 & 1.2 & 14.8 & 0.36 & 1.4 & 6.7 & 3.9 & 16.3 & 0.83 \\
U-Net & multiscale, local & 0.39 & 1.8 & 1.7 & 2.8 & 0.5 & 2.2 & 1.00 & 3.3 & 5.9 & 3.4 & 11.7 & 0.94 \\
ResNet & local, convolution & 0.48 & 4.6 & 4.4 & 7.0 & 1.5 & 7.3 & 0.99 & 3.7 & 9.7 & 3.3 & 14.3 & 0.95 \\
MeshGraphNet & local, message passing & 0.55 & 4.6 & 5.1 & 6.7 & 1.5 & 6.4 & 0.99 & 4.5 & 9.8 & 2.9 & 13.3 & 0.95 \\
PointNet & one global feature & 0.39 & 2.5 & 5.2 & 7.6 & 1.1 & 8.8 & 0.67 & 1.7 & 8.9 & 4.4 & 15.6 & 0.93 \\
MLP & none (pointwise) & 0.42 & 4.4 & 6.0 & 18.1 & 3.1 & 9.6 & 0.85 & 3.2 & 15.9 & 3.7 & 16.2 & 0.86 \\
\hline
$k$NN regression & no learning & 2.07 & 24.1 & 10.2 & 26.2 & 11.1 & 21.7 & 0.93 & -- & 19.4 & 7.3 & 16.7 & 0.85 \\
Sensor interpolation & test-case sensors & 1.50 & 12.7 & 10.4 & 62.9 & 27.5 & 7.5 & 0.52 & -- & 53.7 & 19.6 & 3.8 & 0.56 \\
\end{tabular}
\end{ruledtabular}
\end{table*}

Figure~\ref{fig:bbfields} shows the error fields of all eight architectures on the separated held-out case. The velocity
errors of all architectures are located in the shear layer and the near wake, but differ in magnitude and spatial
distribution: DeepONet spreads its error
along the whole wake, the ResNet and MeshGraphNet concentrate theirs in the near wake, and Transolver has the smallest error
in the pressure field. The pressure errors separate the architectures more clearly. Those that exchange information over the whole
domain or across scales, \ours{}, Transolver and the U-Net, show only a faint ring of error around the leading edge; the local and
pointwise networks leave lobes of error in the outer flow, the same signature as the local-only variant of
Sec.~\ref{sec:spectral}.

Table~\ref{tab:backbones} gives the numbers. Transolver is the most accurate for the field variables ($u$ \TrU\%, $p$ \TrP\%), and at
the wall on the held-out conditions the U-Net is ($C_f$ \UnCf\% against \HoCf\%), reaching \UnBandUA\% in velocity below
$10^{-4}c$ against \HoBandUA\% for \ours{}; on the new airfoils its advantage in skin friction disappears (\UnNewCf\% against
\NewCf\%). The separation point over the 96 database cases is reproduced with $R^2_\mathrm{sep}$ between \BbSepRLo{}
(DeepONet) and 0.99 or more (the U-Net, Transolver, MeshGraphNet and the ResNet), and \HoSepR{} for \ours{}. Every
architecture includes the wall factor, and none approaches the value $R^2_\mathrm{sep}=\NwSepR$ obtained by \ours{} without
it. On the new airfoils at unseen
conditions the architectures lie between \DoNewSepR{} (DeepONet) and \MgNewSepR{} (MeshGraphNet), with \ours{} at \HoNewSepR{}
and the non-learned $k$-nearest-neighbor regression at \KnNewSepR. A field-averaged error does not rank the architectures by
any wall quantity: the MLP has a lower
velocity error than \ours{} and a surface-pressure error seven times larger. Interpolating the sensors of the test case
itself gives a $C_p$ error of 63\%, so the wall quantities are obtained from what the models learn across cases, not from the
observations of the test case.

\subsubsection{Surface-pressure accuracy and computational cost}\label{sec:cost}

\begin{figure*}[!tbp]
\centering
\includegraphics[width=\textwidth]{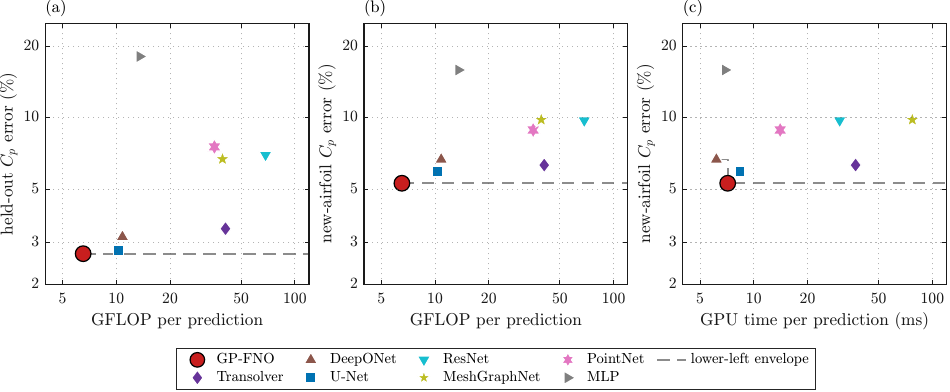}
\caption{\label{fig:cpcost}Surface-pressure accuracy against the cost of one prediction for the eight architectures, which
share the inputs, outputs, wall factor, loss and training of \ours{}. (a) Held-out and (b) new-airfoil $C_p$ error against
the arithmetic of one prediction; (c) new-airfoil $C_p$ error against the GPU time of one prediction. Dashed: lower-left
envelope, the architectures that no other beats in both accuracy and cost.}
\end{figure*}

\begin{table*}[!tp]
\caption{\label{tab:cost}Size and cost of one prediction on the $256\times80$ grid (batch of one, single precision):
real parameters, arithmetic, and wall-clock time on the GPU and on one CPU core (\HwGpu, \HwCpu); training time as
recorded on the training machine (6000 epochs). Last columns: held-out and new-airfoil $C_p$ error, for reference.}
\setlength{\tabcolsep}{2.6pt}
\begin{ruledtabular}
\begin{tabular}{lrrrrrrr}
Model & Param. (M) & GFLOP & GPU (ms) & CPU (s) & Train (min) & $C_p$ (\%) & new $C_p$ (\%) \\
\hline
GP-FNO & 12.72 & 6.5 & 7.1 & 0.21 & 124 & 2.7 & 5.3 \\
\quad global path along $\xi$ only & 0.67 & 5.5 & 5.1 & 0.15 & 91 & 1.9 & 5.8 \\
\quad local path only & 0.14 & 5.1 & 3.7 & 0.10 & 63 & 2.1 & 4.9 \\
Transolver & 0.95 & 40.9 & 37.1 & 1.04 & 571 & 3.4 & 6.3 \\
DeepONet & 0.59 & 10.8 & 6.2 & 0.21 & 64 & 3.2 & 6.7 \\
U-Net & 8.64 & 10.3 & 8.4 & 0.17 & 85 & 2.8 & 5.9 \\
ResNet & 1.67 & 68.5 & 30.3 & 0.83 & 440 & 7.0 & 9.7 \\
MeshGraphNet & 0.36 & 39.3 & 77.2 & 3.22 & 1490 & 6.7 & 9.8 \\
PointNet & 0.87 & 35.4 & 14.1 & 0.61 & 148 & 7.6 & 8.9 \\
MLP & 0.33 & 13.5 & 6.9 & 0.26 & 76 & 18.1 & 15.9 \\
\end{tabular}
\end{ruledtabular}
\end{table*}

Surface pressure determines the lift, and it is the quantity in which the architectures differ most: the models that
exchange information between nodes reach 2.7--7.6\% on the held-out conditions, the MLP 18\%. Figure~\ref{fig:cpcost}
places this accuracy against the cost of a prediction, counted in floating-point operations (FLOP; 1\,GFLOP is $10^9$
FLOP) as the multiply-adds of convolutions, matrix products and einsums (2 FLOP each, complex products of the spectral weights at 8 real FLOP) plus the Fourier transforms at
$2.5N\log_2N$ per real transform of length $N$ (Table~\ref{tab:cost}).

\ours{} occupies the lower-left corner of every panel. In held-out surface pressure it is level with the U-Net (\HoCp\%
against \UnCp\%, a difference smaller than the $\pm$\HoCpPm{} percentage-point variation between the two training seeds of \ours{}), and in new-airfoil
surface pressure it is the most accurate (\NewCp\%, against \UnNewCp\% for the U-Net, \TrNewCp\% for Transolver and
\PnNewCp--\MgNewCp\% for PointNet, the ResNet and MeshGraphNet). It needs the least arithmetic of the eight, \HoGf{} GFLOP per
prediction, against \UnGf{} for the U-Net, \TrGf{} for Transolver and \RnGf{} for the ResNet [Fig.~\ref{fig:cpcost}(a, b)].
No other architecture is both less expensive and more accurate in surface pressure. Within the family of \ours{}, the variant
without the global path (Sec.~\ref{sec:spectral}) reaches the same surface-pressure accuracy at lower cost (\LocCp\% and \LocNewCp\%,
\LocGf{} GFLOP). Its error in the pressure field, however, much larger (\LocP\% against \HoP\%), as is its
skin-friction error on the new airfoils (\LocNewCf\% against \NewCf\%). The benefit of the global path therefore appears in
the pressure field and the wall shear rather than in the surface pressure. Higher computational cost does not translate
into higher accuracy: Transolver, MeshGraphNet, PointNet and the ResNet need five to eleven times the arithmetic of \ours{} and
have larger $C_p$ errors on both sets.

The low cost of \ours{} results from the structure of the Fourier layer. Of the \HoGf{} GFLOP, the local path takes \GfLocalPct\%, the
Fourier transforms \GfFftPct\% and the spectral weights only \GfSpecPct\%: the global path, which carries the pressure
coupling of the whole domain, multiplies only the few hundred retained coefficients of each channel, so its
$1.27\times10^7$ parameters cost memory but almost no time. In wall-clock time [Fig.~\ref{fig:cpcost}(c)] \ours{} takes
\HoMs\,ms on the GPU, within a millisecond of the fastest architectures, DeepONet (\DonMs\,ms) and the MLP (\MlpMs\,ms), which
are less accurate, and four to eleven times faster than Transolver, the ResNet and MeshGraphNet; on one core of the central processing unit (CPU) it takes
\HoCpu\,s, about $1/20\,000$ of the 1.2 core-hours of the Fluent solution of the same condition.

Among architectures that receive the same inputs, biases and training, \ours{} therefore gives surface pressure as
accurate as the best on the held-out conditions and the most accurate on new airfoils, at the least arithmetic per
prediction. Other
architectures lead in other quantities, Transolver in the pressure field and the U-Net in held-out skin friction, and they transfer
differently to finer grids (Sec.~\ref{sec:resolution}). The wall quantities of five architectures on two new airfoils are
compared with OpenFOAM in Appendix~\ref{app:models} (Fig.~\ref{fig:bbsurf}).

%=====================================================================
\section{Conclusions}\label{sec:conclusions}
%=====================================================================

\begin{table*}[!tbp]
\caption{\label{tab:roadmap}Roadmap: each bias of \ours{}, the property of the flow or the numerics that motivates it,
what changes when it is removed or altered, how it acts in the trained model, and where it is examined. Held-out errors
against the reference field; $R^2_\mathrm{sep}$ over the 96 database cases, Eq.~(\ref{eq:r2sep}).}
\renewcommand{\arraystretch}{1.12}
\small
\begin{ruledtabular}
\begin{tabular}{lllll}
Part & Choice, and what motivates it & Measured effect & Mechanism in the trained model & Where \\
\hline
\multicolumn{5}{l}{\emph{Inductive biases}}\\
\quad Grid & \parbox[t]{1.45in}{\raggedright body-fitted O-grid, periodic around the airfoil: the Fourier transform needs a periodic rectangle, and the wall becomes one grid row\strut} & \parbox[t]{1.36in}{\raggedright shared by all models; not removed\strut} & \parbox[t]{1.58in}{\raggedright 44 of the 80 rows within $0.02c$ of the wall\strut} & Sec.~\ref{sec:geometry} \\
\quad Global path & \parbox[t]{1.45in}{\raggedright two-dimensional Fourier layers: pressure obeys a Poisson equation and couples the whole domain\strut} & \parbox[t]{1.36in}{\raggedright removed: $p$ beyond $0.1c$ \HoBandPE\%$\to$\LocBandPE\%, $R^2_\mathrm{sep}$ \HoSepRMean$\to$\LocSepRMean; along $\xi$ only: $C_f$ $\HoCfMean\%\to\XiCfMean\%$; surface $C_p$ unchanged\strut} & \parbox[t]{1.58in}{\raggedright supplies \GlobLBOut--\GlobLDWall\% of each layer's update, most in the last layer near the wall\strut} & Sec.~\ref{sec:spectral} \\
\quad Local path & \parbox[t]{1.45in}{\raggedright $3\times3$ stencils: features finer than the retained modes\strut} & \parbox[t]{1.36in}{\raggedright on $512\times159$: $C_p$ \RfTrainCp\%$\to$\RfBothDirCp\% directly, \RfBothDilCp\% with dilated kernels\strut} & \parbox[t]{1.58in}{\raggedright in the refinement test, the source of the loss of accuracy on finer grids\strut} & Sec.~\ref{sec:resolution} \\
\quad Retained modes & \parbox[t]{1.45in}{\raggedright $32\times12$: hold \SpecCpWall\% of the wall-pressure energy\strut} & \parbox[t]{1.36in}{\raggedright halved: $C_p$ \SwBaseCp\%$\to$\SwMACp\%, $R^2_\mathrm{sep}$ \SwBaseSepR$\to$\SwMASepR{} (1000 epochs)\strut} & \parbox[t]{1.58in}{\raggedright every retained mode used, \WRatioMin--\WRatioMax{} times its initial weight\strut} & Sec.~\ref{sec:capacity} \\
\quad Capacity & \parbox[t]{1.45in}{\raggedright width 64, four layers\strut} & \parbox[t]{1.36in}{\raggedright width 32--128: $C_p$ \SwWACp--\SwWCCp\%; three layers: $R^2_\mathrm{sep}$ \SwDASepR{} (1000 epochs)\strut} & \parbox[t]{1.58in}{\raggedright the field between the sensors converges before the misfit at them (Appendix~\ref{app:training})\strut} & Sec.~\ref{sec:capacity} \\
\quad FiLM & \parbox[t]{1.45in}{\raggedright $(Re,\alpha)$ act on the whole flow and modulate every layer\strut} & \parbox[t]{1.36in}{\raggedright removed: no loss inside the ranges; lift unbounded in extrapolation\strut} & \parbox[t]{1.58in}{\raggedright a nearly constant gain, varying by at most \GamVarA{} inside the ranges\strut} & Sec.~\ref{sec:film} \\[2pt]
\quad Wall factor & \parbox[t]{1.45in}{\raggedright $\mathbf u=r(d)\hat{\mathbf u}$: exact no-slip; $\tau_w$ becomes an output value, Eq.~(\ref{eq:wallfactor})\strut} & \parbox[t]{1.36in}{\raggedright removed: $C_f$ \HoCf\%$\to$\NwCf\%, $R^2_\mathrm{sep}$ \HoSepR$\to$\NwSepR\strut} & \parbox[t]{1.58in}{\raggedright without it the first-row error has \RippleHigh{} times the energy beyond the retained modes\strut} & Sec.~\ref{sec:wall} \\
\multicolumn{5}{l}{\emph{Learning biases}}\\
\quad Continuity & \parbox[t]{1.45in}{\raggedright relative residual, $\lambda=0.1$: mass is conserved at the 95\% of nodes without data, and the data satisfy it\strut} & \parbox[t]{1.36in}{\raggedright removed: memorization ratio \HoMemo$\to$\NcMemo\strut} & \parbox[t]{1.58in}{\raggedright without it the error does not fall away from the sensors, and the median residual is \ContRatio{} times that of \ours{}\strut} & Sec.~\ref{sec:continuity} \\
\quad Momentum & \parbox[t]{1.45in}{\raggedright optional, on $d\ge10^{-3}c$ only, where the data satisfy it\strut} & \parbox[t]{1.36in}{\raggedright added: $C_p$ \HoCp\%$\to$\MbCp--\MaCp\%, $C_f$ \HoCf\%$\to$\MaCf--\McCf\%\strut} & \parbox[t]{1.58in}{\raggedright lowers the pressure error at every wall distance; first rows untouched\strut} & Sec.~\ref{sec:momentum} \\[2pt]
\multicolumn{5}{l}{\emph{Observational bias}}\\
\quad Sensors & \parbox[t]{1.45in}{\raggedright fixed 5\% per case, stratified by geometry and $\alpha$ only\strut} & \parbox[t]{1.36in}{\raggedright 4096 of 81\,920 values per case\strut} & \parbox[t]{1.58in}{\raggedright \ShareThree\% of the nodes within three nodes of a sensor\strut} & Secs.~\ref{sec:sparse}, \ref{sec:continuity} \\
\end{tabular}
\end{ruledtabular}
\end{table*}

We developed a body-fitted, geometry- and physics-encoded Fourier neural operator, \ours{}, that learns turbulent flow
over airfoils, computed with the RANS equations, from training solutions observed at 5\% of their nodes, and predicts new
flow conditions and new airfoils without any observation of them. Under a single protocol, each of its inductive and learning
biases was removed or altered in turn, and the effect was quantified in the flow fields, the wall quantities and the trained
network. The main conclusions are as follows.

(1) With every bias in place, \ours{} reproduces the held-out surface pressure, lift and skin friction, which is never
supervised, to \HoCp\%, \HoCl\% and \HoCf\%, and the separation point with $R^2_\mathrm{sep}=\HoSepR$ over the 96 database
cases. Trained on solutions of one RANS solver and evaluated against another, it reaches median errors on \Nnew{} solutions of eight new airfoils
(\NewCp\% in $C_p$, \NewCl\% in $C_L$, \NewCf\% in $C_f$) close to the difference between the two codes themselves, which
it tracks airfoil by airfoil, with a correlation of \GapCorr{} in $C_p$.

(2) Among the inductive biases, the wall factor, which builds no-slip into the output, decides whether skin friction and
separation can be learned: without it the skin-friction error rises fivefold and $R^2_\mathrm{sep}$ falls from \HoSepR{} to
\NwSepR. The two-dimensional global path carries the pressure coupling of the volume and makes the wall shear consistent with
it, and the retained modes are best chosen from the wall spectrum of the data. On refined grids, the loss of accuracy we
observed came from the local path, and dilating its kernels recovered it, a gain in consistency rather than in resolution.
Parameter conditioning only bounds extrapolation, and width and depth matter little inside the data.

(3) Among the learning biases, the continuity residual is the only constraint on the unobserved nodes; without it the network
memorizes its sensors far more, with a memorization ratio of \NcMemo{} instead of \HoMemo, and its held-out field errors rise. The momentum residual, admissible only away
from the wall, improves the pressure everywhere but not the wall shear.

(4) Against seven other architectures trained with the same inputs, biases and data, \ours{} is as accurate in surface
pressure as the best on the held-out conditions, \HoCp\%, and the most accurate on the new airfoils, \NewCp\%, at the least
arithmetic, \HoGf{} GFLOP per
prediction, a sixth of that of Transolver and a tenth of that of the ResNet, and in \HoMs\,ms on a GPU.

Together these results indicate that surrogates that are accurate under sparse supervision and inexpensive to evaluate
are obtained not from larger or more elaborate networks, but from architectures whose biases are matched to the flow. Each
bias of \ours{} addresses a specific property of the flow or of the numerics, and each was tested by removing it. These
properties are the dependence of the pressure on the flow in the whole domain, expressed by its Poisson equation, the
separation of scales at the wall, and the conservation of mass at the unobserved nodes.
Table~\ref{tab:roadmap} condenses the investigation into a roadmap that states, for every bias, the property that motivates
it, its measured effect and how it acts in the trained model. We offer it, together with the protocol of
Sec.~\ref{sec:protocol} and the reporting of error fields and derived wall quantities rather than field-averaged errors
alone, as a basis on which researchers can design and test sparse-data operators for other wall-bounded flows.

This study has several limitations. The three training airfoils span a narrow range of camber and
thickness, and the data barely extend beyond the onset of stall. All solutions are steady,
incompressible RANS solutions with a single turbulence closure and no transition model, and the
skin-friction target differs from the solver's wall shear by a median of \CfTarget\%. The body-fitted
grid was not removed, each comparison architecture was trained with one seed, the sensor fraction was
fixed at 5\%, and $\delta$, $\lambda$ and $\kappa$ were set a priori rather than swept. Because every
model was kept at its best held-out state, the held-out conditions are not fully independent of
training; the new airfoils are.

The comparison also raises an open question. Every architecture that carried the biases of \ours{}
predicted the separation point of the database cases with $R^2_\mathrm{sep}\ge\BbSepRLo$, whereas
\ours{} without the wall factor reached \NwSepR, suggesting that for the wall quantities the biases
matter more than the backbone. Testing this requires a factorial study that also removes the wall
factor and the continuity residual from the other backbones. Future work should also address shapes
that vary camber and thickness independently, data beyond stall, a range of sensor fractions,
three-dimensional flows, and unsteady flows such as dynamic stall or vortex shedding, in which the
operator must also predict the temporal evolution of the flow.

\raggedbottom
\begin{acknowledgments}
The authors thank the Indian Institute of Technology Gandhinagar for computational resources.
\end{acknowledgments}

\section*{Author Declarations}
\subsection*{Conflict of Interest}
The authors have no conflicts to disclose.

\subsection*{Author Contributions}
\textbf{Ada Sinha}: Conceptualization (equal); Formal analysis (lead); Investigation (equal); Methodology (equal);
Software (lead); Visualization (equal); Writing -- original draft (equal).
\textbf{Kushagreek Basu}: Conceptualization (equal); Data curation (lead); Investigation (equal); Methodology (equal);
Validation (lead); Visualization (equal); Writing -- original draft (equal); Writing -- review \& editing (equal).
\textbf{Shubhangi Bansude}: Conceptualization (equal); Project administration (lead);
Resources (lead); Supervision (lead); Writing -- review \& editing (equal).

\section*{Data Availability}
The data that support the findings of this study are available from the corresponding author upon reasonable request.

\FloatBarrier
\appendix
\begin{figure*}[!tbp]
\centering
\includegraphics[width=\textwidth]{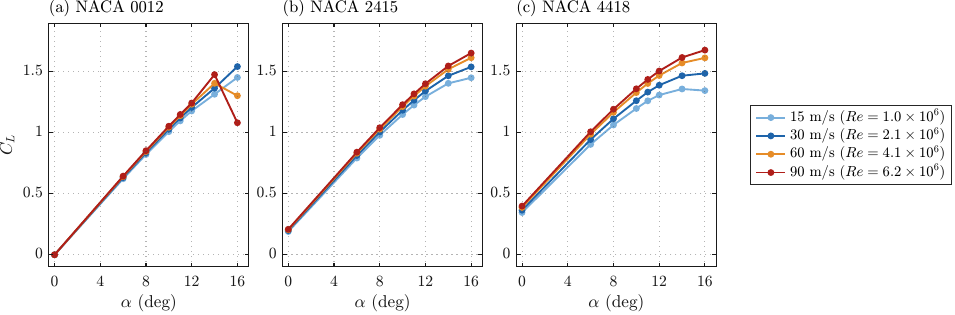}
\caption{\label{fig:dblift}Lift of the 96 database solutions, one panel per training airfoil, at the four speeds.}
\end{figure*}

\begin{figure*}[!tbp]
\centering
\includegraphics[width=\textwidth]{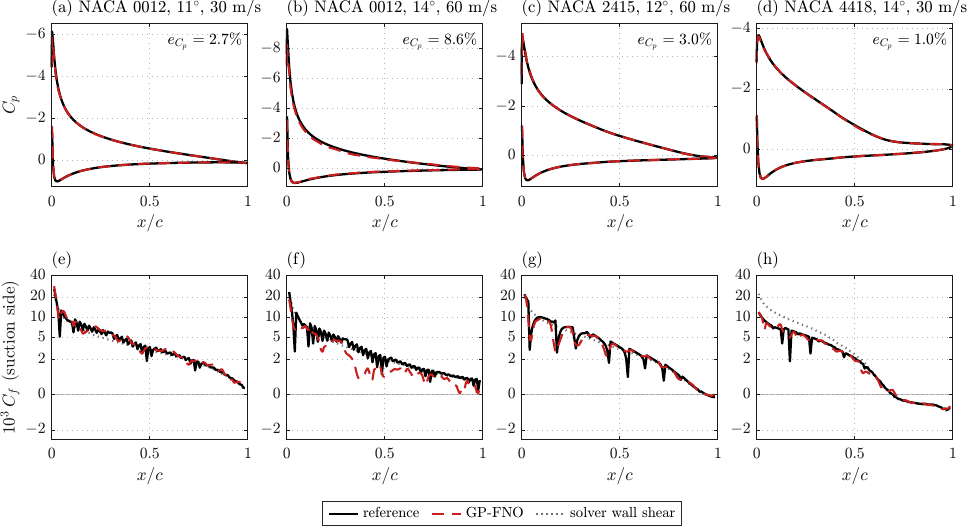}
\caption{\label{fig:cftarget}Derived and solver wall shear on four held-out conditions, with the prediction of \ours{}:
(a--d) surface pressure, with the relative error of \ours{}; (e--h) suction-side skin friction on a symmetric logarithmic
axis (linear for $|10^3C_f|<1$), derived from the reference field (reference) and as the solver's own wall shear.}
\end{figure*}

\begin{table*}[!t]
\caption{\label{tab:heldout}The complete operator on the eight held-out conditions, against the reference field. Errors are
area-weighted relative errors of the field, Eq.~(\ref{eq:l2}), the relative error of the surface pressure, and the median
relative error of the derived wall shear, all in percent; $x_\mathrm{sep}$ is the first sign change of the suction-side
wall shear (--: attached).}
\begin{ruledtabular}
\begin{tabular}{lrrrrrrrrrr}
 & & & \multicolumn{4}{c}{Error (\%)} & \multicolumn{2}{c}{$C_L$} & \multicolumn{2}{c}{$x_\mathrm{sep}/c$} \\
\cline{4-7}\cline{8-9}\cline{10-11}
Airfoil & $\alpha$ ($^\circ$) & $U_\infty$ (m/s) & $u$ & $p$ & $C_p$ & $C_f$ & ref. & GP-FNO & ref. & GP-FNO \\
\hline
NACA 0012 & 6 & 60 & 0.27 & 2.3 & 2.1 & 12.1 & 0.641 & 0.639 & -- & -- \\
NACA 0012 & 11 & 30 & 0.38 & 1.1 & 2.7 & 7.9 & 1.122 & 1.115 & -- & -- \\
NACA 0012 & 14 & 60 & 1.32 & 5.7 & 8.6 & 21.1 & 1.407 & 1.321 & -- & -- \\
NACA 2415 & 8 & 30 & 0.27 & 1.3 & 1.8 & 3.9 & 1.007 & 1.006 & -- & -- \\
NACA 2415 & 12 & 60 & 0.46 & 1.2 & 3.0 & 5.8 & 1.383 & 1.385 & 0.95 & 0.96 \\
NACA 4418 & 6 & 30 & 0.40 & 1.4 & 1.2 & 6.0 & 0.945 & 0.939 & 0.93 & 0.94 \\
NACA 4418 & 11 & 60 & 0.40 & 1.1 & 1.1 & 6.0 & 1.405 & 1.405 & 0.87 & 0.86 \\
NACA 4418 & 14 & 30 & 0.50 & 1.4 & 1.0 & 6.4 & 1.469 & 1.462 & 0.70 & 0.71 \\
\end{tabular}
\end{ruledtabular}
\end{table*}

\begin{figure*}[!tbp]
\centering
\includegraphics[width=0.95\textwidth]{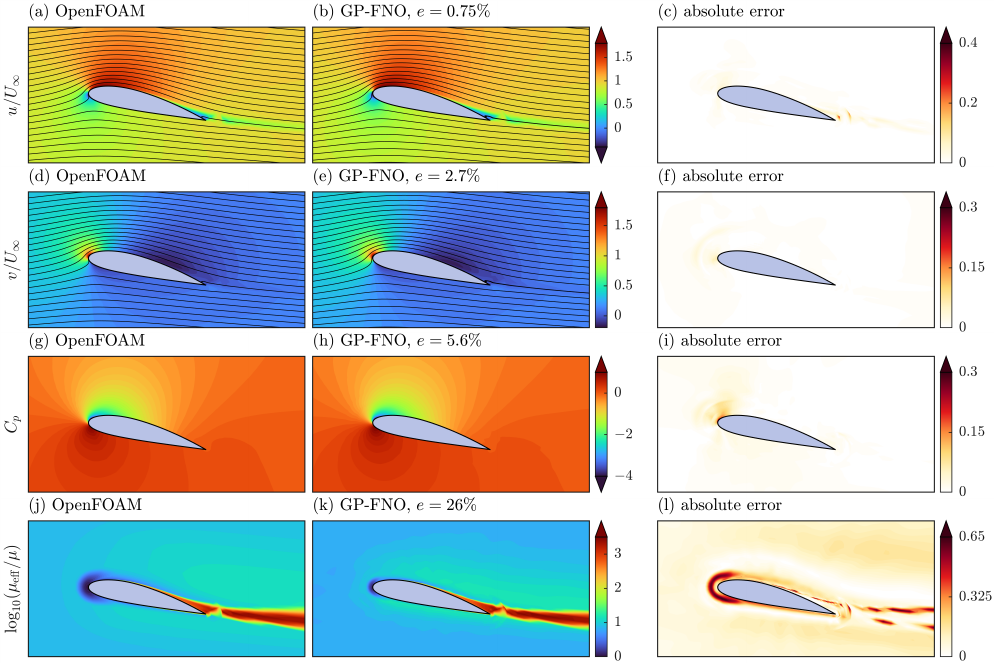}
\caption{\label{fig:newhighspeed}As Fig.~\ref{fig:newstall}, for NACA~3417, $\alpha=13^\circ$, 75\,m/s.}
\end{figure*}

\begin{figure*}[!tbp]
\centering
\includegraphics[width=0.95\textwidth]{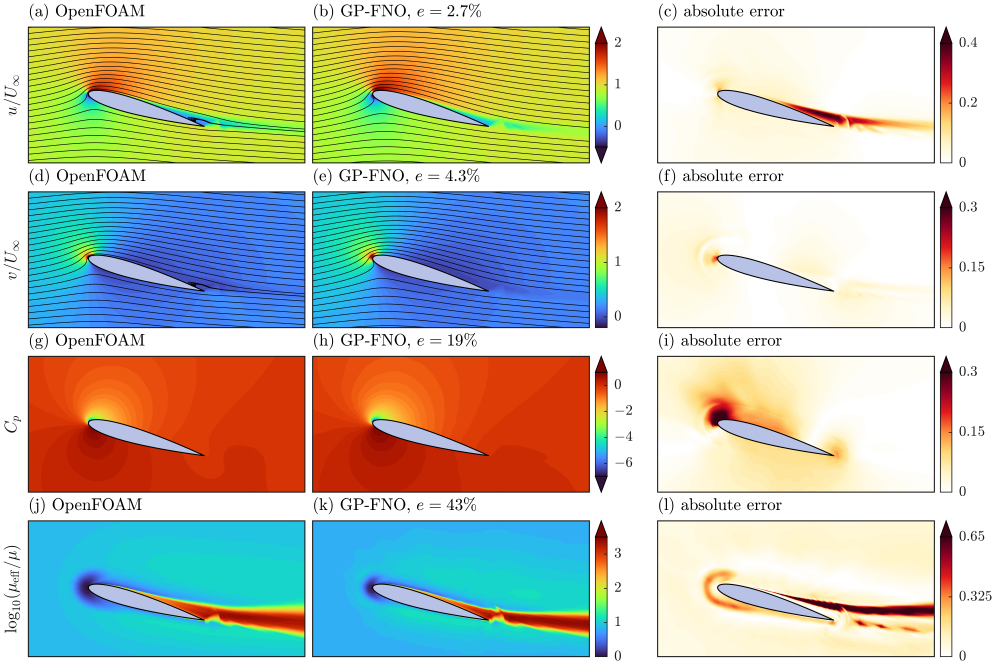}
\caption{\label{fig:newlargest}As Fig.~\ref{fig:newstall}, for NACA~0013, $\alpha=16^\circ$, 30\,m/s, a new airfoil at a
trained condition at the edge of the angle range. The error panels saturate, as the arrows on the color bars indicate.}
\end{figure*}

\begin{figure*}[!tbp]
\centering
\includegraphics[width=\textwidth]{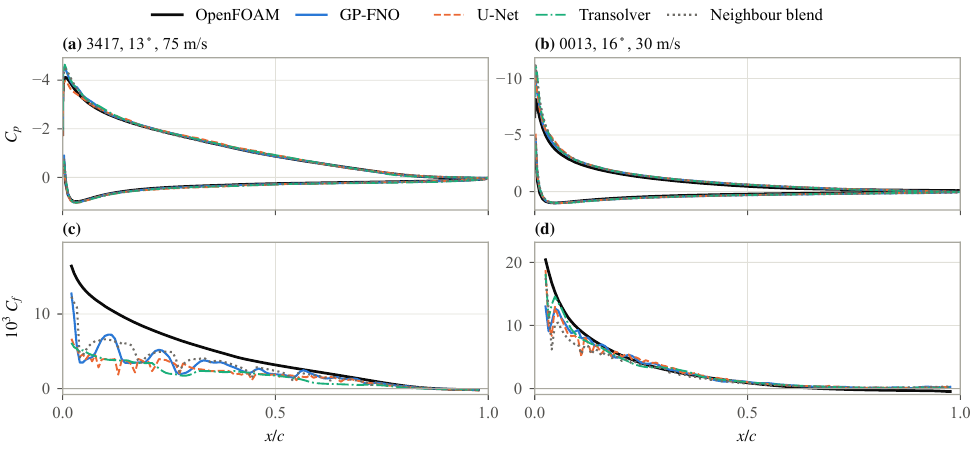}
\caption{\label{fig:newsurfextra}Surface pressure and suction-side skin friction of the cases of Figs.~\ref{fig:newhighspeed}
and \ref{fig:newlargest}, as in Fig.~\ref{fig:newsurf}.}
\end{figure*}

\begin{table*}[!tp]
\caption{\label{tab:sweep}Width, retained modes ($m_\xi\times m_\eta$, so that $-m_\xi\le k_\xi<m_\xi$ and $0\le k_\eta<m_\eta$)
and depth, each trained for 1000 epochs under the same protocol and scored on the held-out conditions;
$R^2_\mathrm{sep}$: coefficient of determination of the separation point over the 96 database cases.}
\setlength{\tabcolsep}{2.4pt}
\begin{ruledtabular}
\begin{tabular}{rrrrrrrrrrrl}
Width & Modes & Layers & Param. (M) & GFLOP & $p$ (\%) & $C_p$ (\%) & $C_L$ (\%) & $C_f$ (\%) & $R^2_\mathrm{sep}$ & $M$ & \\
\hline
64 & 8\,$\times$\,6 & 4 & 3.3 & 6.5 & 3.2 & 3.6 & 1.59 & 7.5 & $-$0.28 & 2.4 &  \\
64 & 16\,$\times$\,12 & 4 & 12.7 & 6.5 & 2.4 & 2.5 & 0.85 & 8.9 & 0.22 & 4.7 & \textbf{adopted} \\
64 & 24\,$\times$\,18 & 4 & 28.5 & 6.6 & 2.6 & 2.3 & 1.01 & 8.0 & 0.13 & 6.1 &  \\
32 & 16\,$\times$\,12 & 4 & 3.2 & 1.8 & 3.3 & 2.5 & 0.83 & 9.4 & 0.61 & 2.9 &  \\
128 & 16\,$\times$\,12 & 4 & 50.8 & 24.4 & 2.4 & 2.5 & 0.94 & 7.6 & 0.50 & 6.2 &  \\
64 & 16\,$\times$\,12 & 3 & 9.5 & 4.6 & 2.8 & 2.5 & 0.86 & 8.3 & $-$2.11 & 3.2 &  \\
64 & 16\,$\times$\,12 & 6 & 19.1 & 10.4 & 2.4 & 2.3 & 0.58 & 7.2 & 0.64 & 6.0 &  \\
\end{tabular}
\end{ruledtabular}
\end{table*}

\section{Database details}\label{app:data}

The Fluent solutions use domains that extend 5--$7.5c$ upstream and to the sides and 10--$15c$ downstream, a velocity inlet
(turbulence intensity 5\%, viscosity ratio 10) and a pressure outlet, and ran for 2000--6000 iterations. In 87 of the 96
solutions the force coefficients change by less than 1\% over the last 500 iterations; the other nine, at $14$--$16^\circ$,
keep a slow oscillation, and their last iterate is used. The free-stream Mach number is 0.04--0.26 and the density is
constant. The local Mach number exceeds 0.3 somewhere in 44 of the 96 solutions, all 24 at 90\,m/s, 19 of the 24 at
60\,m/s and one at 30\,m/s, over at most 18\% of the exported nodes of a case, and it reaches 0.86 at the strongest suction
peak; in the nine solutions at 90\,m/s where it exceeds 0.6, the data describe the incompressible model problem rather than
the physical flow. No transition model is used, so the boundary layer is turbulent
from the leading edge. Figure~\ref{fig:dblift} shows the lift of all 96 solutions. Maximum lift is passed inside the
$16^\circ$ range only for NACA~0012 at 60 and 90\,m/s and NACA~4418 at 15\,m/s. Over the 96 cases the derived reference shear differs from the solver's own wall shear by a median of
\CfTarget\% (10th to 90th percentile \CfTargetLo--\CfTargetHi\%).

Figure~\ref{fig:cftarget} sets the derived reference shear beside the solver's wall shear on four held-out conditions. On
three of them the two nearly coincide; on NACA~4418 at $14^\circ$ and 30\,m/s the derived shear lies below the solver's
ahead of mid-chord [Fig.~\ref{fig:cftarget}(h)], and \ours{} follows the derived target on which it was trained.

\section{OpenFOAM solutions and their transfer to the grid}\label{app:openfoam}

The OpenFOAM solutions ran for 10\,000 iterations with the inlet turbulence of the training data. Each solution is classified from its lift
history over the last 2000 iterations as steady (range below 1\% and drift between the two halves of the window below 0.2\%)
or oscillating; of the \Nnew{} new-airfoil solutions, \StN{} are steady and \OsN{} are not, among them every case at
$16^\circ$, and the lift of every solution varies by less than 0.06 over the window. Fields and wall shear are averaged over
this window. The median $C_p$ error of \ours{} is \StCp\% on the steady solutions and \OsCp\% on the oscillating ones.

The transfer to the O-grid was checked against OpenFOAM's own wall values. Near the wall, interpolation uses arc length and
the coordinate $\operatorname{arcsinh}(d/d_0)$ with $d_0=8\times10^{-5}c$, which is linear next to the wall and logarithmic
away from it. A purely logarithmic coordinate would overestimate the velocity between widely spaced cells next to the wall,
and the bias of the derived shear would reach 12.8\% at 20\,m/s; the adopted coordinate reduces it to 0.35\%. The wall row
takes OpenFOAM's wall pressure directly, mapped along the arc length. On the held-out conditions OpenFOAM's $C_p$ sits above
the training data by a nearly uniform offset of 0.008--0.029, which affects no force; this is the offset that
Eq.~(\ref{eq:gauge}) removes.

\section{Held-out cases}\label{app:cases}

Table~\ref{tab:heldout} lists the complete operator on each held-out condition.

\section{Supplementary results}\label{app:extra}

Figures~\ref{fig:newhighspeed} and \ref{fig:newlargest} add two new-airfoil cases to the one of Sec.~\ref{sec:res_new},
and Fig.~\ref{fig:newsurfextra} gives their wall quantities. NACA~3417 at $13^\circ$ and 75\,m/s is reproduced closely in
its fields, but its skin friction near the leading edge falls short of OpenFOAM's, with a $C_f$ error of 30.7\%, and every
learned model and the neighbour blend share this underprediction [Fig.~\ref{fig:newsurfextra}(c)]. NACA~0013 at $16^\circ$ and 30\,m/s lies at the edge of the angle range,
on the symmetric airfoil whose training neighbor has the largest code gap, and it has the second-largest surface-pressure
error of all new-airfoil solutions (16.8\% in $C_p$, 15.0\% in lift): \ours{} overshoots the suction peak and the separated
region, and the velocity error of this case, 2.7\%, is the largest of the cases shown.

Table~\ref{tab:sweep} lists the sweep of retained modes, width and depth of Sec.~\ref{sec:capacity}. The retained modes
change the parameter count nearly ninefold but barely change the arithmetic (\SwMAGf{} against \SwMCGf{} GFLOP), because the
spectral weights multiply only the retained coefficients; width changes both.

\FloatBarrier
\section{Training}\label{app:training}

The lifting $P$ is a pointwise linear layer from 7 to 64 channels, and the decoder $Q$ is pointwise (64--128--4). The
spectral weights are initialized with independent uniform real and imaginary parts on $[0,1/64^2)$, the level marked in
Fig.~\ref{fig:capacity}(d). Gradients are clipped at norm 1. All models were trained, and all checkpoints saved, with PyTorch on one
NVIDIA GeForce RTX 5050 GPU (8\,GB); the evaluation and the timings of Table~\ref{tab:cost} were run on a second machine
(\HwGpu, \HwCpu).

\begin{figure}[H]
\centering
\includegraphics[width=\columnwidth]{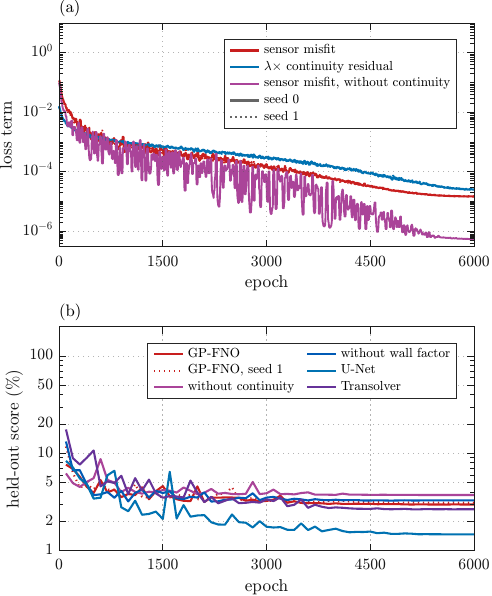}
\caption{\label{fig:history}Training. (a) Sensor misfit and weighted continuity residual of \ours{}, Eq.~(\ref{eq:loss}),
for two seeds, and the sensor misfit of the variant without the continuity residual (running median over 25 epochs).
(b) Held-out score, the mean area-weighted error of $u$, $v$ and $p$ and of the near-wall velocity, recorded every 100
epochs.}
\end{figure}

Figure~\ref{fig:history} shows the training of \ours{}. The two terms of Eq.~(\ref{eq:loss}) fall
together, and at the end they are comparable, as $\lambda=0.1$ intends. Without the continuity term the sensor misfit ends
27 times lower, the memorization of Sec.~\ref{sec:continuity} seen from the loss. The held-out score reaches a plateau near
3\% after about 3000 epochs while the sensor misfit keeps falling: fitting the sensors and learning the field between them
are different tasks. The predictions and metrics of this paper were recomputed from the archived checkpoints with the
original evaluation code and reproduce the archived predictions to a relative difference of $10^{-6}$.

\section{Details of the comparison architectures}\label{app:models}

All comparison architectures act on the same seven inputs, plus $\log_{10}d$, which pointwise models need in order to tell the region next to the wall from the outer boundary layer, and all return the same four raw outputs, to which the wall factor is applied.
The MLP has six hidden layers of width 256. PointNet\cite{qi2017} adds one global feature obtained by max-pooling over the
domain. The ResNet\cite{he2016} has ten residual blocks of width 96 and a $41\times41$ receptive field. The
U-Net\cite{ronneberger2015} has base width 32 and four levels, down to $16\times5$ nodes, with padding circular in $\xi$.
MeshGraphNet\cite{pfaff2021} passes 12 rounds of messages over the grid graph, with edge features from the physical
geometry. DeepONet\cite{lu2021deeponet} expands each output in 128 trunk functions, with coefficients from a branch network of
$(Re,\alpha)$ and the resampled wall coordinates. Transolver\cite{wu2024transolver} attends among 32 learned slices in each
of five layers. The $k$-nearest-neighbor regression uses the 16 nearest training sensors in the space of inputs. Training
took about one hour for DeepONet and the MLP, \HoTrainMin\,min for \ours{}, \TrTrainMin\,min for Transolver and
\MgTrainMin\,min for MeshGraphNet.

\begin{figure*}[!t]
\centering
\includegraphics[width=0.94\textwidth]{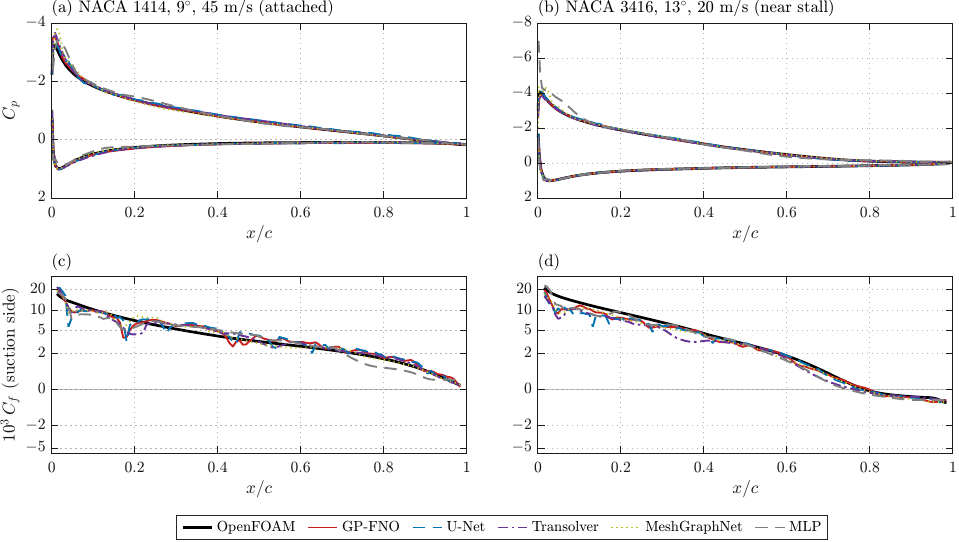}
\caption{\label{fig:bbsurf}Wall quantities of five architectures on two new airfoils at unseen conditions, against
OpenFOAM: (a, b) surface pressure, on the OpenFOAM gauge; (c, d) suction-side skin friction (symmetric logarithmic axis),
with OpenFOAM's own wall shear.}
\end{figure*}

Figure~\ref{fig:bbsurf} compares the wall quantities of five architectures on two new airfoils. In surface pressure all
models but the MLP are close to OpenFOAM. In skin friction the models agree more closely with one another than with
OpenFOAM: all are trained on the same derived target and evaluated against a second solver, so their common deviation is attributable to
the target and to the code gap, not to one architecture.

\bibliography{references_v8}

\end{document}